\documentclass[12pt]{article}
\usepackage{amsmath,amssymb,amsfonts,epsfig,graphicx}

\newcommand{\bse}{\begin{subequations}}
\newcommand{\ese}{\end{subequations}}
\newcommand{\be}{\begin{equation}}
\newcommand{\ee}{\end{equation}}
\newcommand{\bea}{\begin{eqnarray}}
\newcommand{\eea}{\end{eqnarray}}
\newcommand{\ba}{\begin{array}}
\newcommand{\ea}{\end{array}}
\newcommand{\nn}{\nonumber}
\makeatletter \@addtoreset{equation}{section}

\makeatother
\expandafter\ifx\csname mathbbm\endcsname\relax

\else

\fi
\def\by{\times}
\def\1{{\bf 1_{2^k\times2^k}}}
\def\l{\left}
\def\r{\right}
\def\bl{\biggl}
\def\br{\biggr}

\def\N{{\cal N}}

\def\F{{\cal F}}
\def\G{{\cal G}}
\def\L5{{\cal L}_5}
\def\Lodd{{\cal L}_{2k+1}}
\def\M{{\cal M}}

\def\C{\mathbb C}
\def\Tr{{\rm {Tr}}}

\def\pl{plane-wave\ }
\def\lc{light-cone\ }
\def\Ham{Hamiltonian\ }
\def\bg{background}

\def\super{$PSU(2|2)\times PSU(2|2)\times U(1)$}
\begin{document}
\baselineskip 18pt%

\begin{titlepage}
\vspace*{1mm}%
\hfill%
\vbox{
    \halign{#\hfil        \cr
           IPM/P-2004/075 \cr
           hep-th/0501001 \cr
           } 
      }  
\vspace*{10mm}%

\begin{center}
{\Large {\bf Classification of All 1/2 BPS Solutions of\\
the Tiny Graviton Matrix Theory}}%
\vspace*{10mm}

{\bf M. M. Sheikh-Jabbari$^{1,2}$, M. Torabian$^{1,2,3}$}%

\vspace*{0.4cm}
{\it {$^1$Institute for Studies in Theoretical Physics and Mathematics (IPM)\\
P.O.Box 19395-5531, Tehran, IRAN\\
$^2$the Abdus Salam International Centre for Theoretical Physics\\
34014, Trieste, ITALY\\
$^3$Department of Physics, Sharif University of Technology\\
P.O.Box 11365-9161, Tehran, IRAN}}\\
{E-mails: {\tt jabbari, mahdi@theory.ipm.ac.ir}}%
\vspace*{1.5cm}
\end{center}

\begin{center}{\bf Abstract}\end{center}
\begin{quote}
The tiny graviton Matrix theory \cite{tiny} is proposed to
describe DLCQ of type IIB string theory on the maximally
supersymmetric plane-wave or $AdS_5\times S^5$ background. In this
paper we provide further evidence in support of the tiny graviton
conjecture by focusing on the zero energy, half BPS configurations
of this matrix theory and classify all of them. These vacua are
generically of the form of various three sphere giant gravitons.
We clarify the connection between our solutions and the half BPS
configuration in $\N=4$ SYM theory and their gravity duals.
Moreover, using our half BPS solutions,  we show how the tiny
graviton Matrix theory and the mass deformed $D=3,\ \N=8$
superconformal field theories are related to each other.
\end{quote}

\end{titlepage}

\section{Introduction}
Soon after the introduction of (flat) D-branes as dynamical
objects in string theory \cite{Polchinski} it was realized that
the theory residing on $N$ coincident $Dp$-branes is a $p+1$
dimensional $U(N)$ supersymmetric Yang-Mills (SYM) theory with 16
supercharges \cite{Witten}. The above facts have been in the core
of the most interesting developments in string/M- theory in the
past eight years, the BFSS Matrix Theory \cite{BFSS} and the
AdS/CFT correspondence \cite{AdS/CFT}. In the both examples
certain $\alpha'\to 0$ limit of a background with $N$ D-branes
were used to argue for (or obtain) a non-perturbative description
of string/M- theory. In the BFSS case, this was $N$ $D0$-branes,
{\it i.e.} a $U(N)$ $0+1$ dimensional SYM, which was proposed to
describe Discrete Light-Cone Quantization (DLCQ) of M-theory in
the sector with $N$ units of light-cone momentum \cite{BFSS}. In
the AdS/CFT, however, the near horizon limit of the geometry with
$N$ $D3$-branes, {\it i.e.} the $AdS_5\times S^5$ background with
$N$ units of the five-form flux through the $S^5$, was proposed to
be dual (or equivalent) to the ${\cal N}=4,\ D=4$ $U(N)$ SYM. In
another point of view, the latter is the holographic description
of string theory on $AdS_5\times S^5$, as the causal boundary of
the $AdS_5$ geometry is $R\times S^3$ \cite{holography}. The two
conjectures, the BFSS matrix model and the AdS/CFT, have passed
many non-trivial crucial tests and a large class of such checks
are based on analysis of supersymmetric, BPS configurations. BPS
states provide a powerful tool for checking the conjectures
because they are protected against corrections which are often
times out of control in the desired regimes.

The type IIB string $\sigma$-model on the $AdS_5\times S^5$
background has turned out to be very hard to solve, e.g. see
\cite{BPR}, and consequently many tests of the AdS/CFT in string
theory side has been limited to the supergravity limit
(corresponding to large $N$ limit in the dual gauge theory). In a
quest for pushing the duality beyond the supergravity limit it was
shown that the Penrose limit, after which the $AdS_5\times S^5$
geometry goes over to the plane-wave background \cite{BMN, Blau},
opens the possibility of exploring a region where the gauge and
string theories are both perturbatively accessible (for reviews
see \cite{review, Plefka-lectures}).

The ten dimensional, maximally supersymmetric plane-wave
background (here we follow conventions and notations of
\cite{review}.):
\begin{subequations}\label{background}
\begin{align}
 ds^2 = -2 dX^+ dX^- & -\mu^2(X^i X^i + X^a X^a) {(dX^+)}^2 + dX^i dX^i+ dX^a dX^a \\
 F_{+ijkl} &= \frac{4}{g_s} \mu\ \epsilon_{ijkl} \qquad,\qquad
 F_{+abcd}=\frac{4}{g_s} \mu \ \epsilon_{abcd} \\
 &\qquad\quad e^\phi = g_s = \rm{constant}
\end{align}
\end{subequations}%
where $i,a=1,2,3,4$ and $F_5$ is the (self-dual) fiveform field
strength, has a one dimensional {\it light-like} causal boundary
and this leads one to the question whether strings on the
plane-wave background has a holographic description. Such a
theory, if exists, should then be a $0+1$ dimensional (presumably
gauge) Matrix theory.

In \cite{tiny}, through a study of the three-brane giant gravitons
\cite{MST} and their quantization, a $0+1$ dimensional $U(J)$
gauge theory was obtained, the Tiny Graviton Matrix Theory (TGMT).
According to the tiny graviton conjecture, TGMT describes  DLCQ of
type IIB string theory on the plane-wave \eqref{background}, in
the sector with $J$ units of light-cone momentum. Furthermore, it
was  argued that the same theory should also describe DLCQ of type
IIB strings on the $AdS_5\times S^5$ background.

Some pieces of evidence in support of the TGMT conjecture was
presented in \cite{tiny}. In this work we provide further
supportive evidence through a detailed and exhaustive study of all
the 1/2 BPS configurations of the TGMT. For that, in section
\ref{review-section}, we review the statement of the conjecture
and the TGMT Hamiltonian. In section \ref{Zero-Energy-section}, we
focus on the zero energy configurations of the matrix theory and
show that these are generically of the form of concentric fuzzy
three spheres residing in the $X^i$ and/or $X^a$ directions ({\it
cf.} \eqref{background}). In the large $J$ (string theory) limit
these fuzzy spheres become spherical three-brane giants. In
section \ref{SYM-gravity-relation}, we argue how our zero energy
configurations are related to the similar 1/2 BPS configurations
in the type IIB supergravity recently studied in \cite{LLM} and in
the ${\cal N}=4, D=4$ $U(N)$ gauge theory \cite{Berenstein}. In
section \ref{mass-deformed}, we show how the TGMT and the mass
deformed ${\cal N}=8,\ D=3$ SCFT are related to each other.
Explicitly, we argue that the three fuzzy spheres of TGMT are
indeed the quantized (longitudinal) M5-branes of the latter. This
would also shed light on some less clear part of the TGMT, namely
the matrix ${\cal L}_5$ ({\it cf.} the arguments of \cite{tiny} or
section \ref{review-section}). In section
\ref{discussion-section}, we give a summary of our results and an
outlook. Some technical points have been gathered in the
Appendices. In Appendix \ref{convention}, we present our
conventions for the Dirac gamma matrices and some useful
identities. In Appendix \ref{appendixB}, we review the new
construction for the fuzzy spheres presented in \cite{tiny}. This
construction is based on the quantization of Nambu brackets. We
solve the two equations defining a generic fuzzy sphere by {\it
embedding} the fuzzy sphere in a higher dimensional noncommutative
Moyal plane and work out the details of this solution for the
cases of fuzzy two and four spheres. This constitutes a new
construction for the fuzzy spheres, in particular fuzzy three and
four spheres. In the Appendix \ref{superalgebra}, we have
presented the (dynamical part of the) superalgebra of the tiny
graviton matrix theory, namely \super\ algebra and its
representation in terms of matrices.


\section{Review of The Tiny Graviton Matrix Theory}\label{review-section}%
In this section we briefly review the basics of the tiny graviton
conjecture. This is essentially a short summary of \cite{tiny}.

\subsection{The Giant, The Normal, The Tiny}\label{giant-normal-tiny}

As discussed in the introduction, the key objects in both BFSS
matrix theory and the AdS/CFT correspondence are  flat 1/2 BPS
D-branes. In the non-flat backgrounds, where a form flux is turned
on, it is possible to construct 1/2 BPS (topologically) spherical
branes, the giant gravitons \cite{MST}. In order to stabilize a
spherical brane at a finite size we need to exert a repulsive
force on the brane to overcome its tension. This force can be
provided, noting that a spherical $p$-brane carries an (electric)
dipole moment of the $p+1$-form and if we have a $p+2$-form
(magnetic) flux in the background a moving (rotating) brane would
feel a repulsive force. The situation can be arranged such that
the tension and the form-field forces cancel each other. Indeed
this is only possible if the brane is following a light-like
geodesic, in this respect it behaves like any other (super)gravity
mode, hence it was called a giant graviton \cite{MST}. (Note that
the spherical $p$-brane cannot carry $p+1$-form charge, unlike a
flat brane.)

{}From the above argument it follows that the size of the giant is
related to its angular momentum. In the  $AdS_5\times S^5$ or the
plane-wave background, where we have a fiveform flux in the
background, we can stabilize an $S^3$ giant. The size of the giant
and its angular momentum $J$ are related as \cite{MST}%
\be\label{giant-size}%
\left(\frac{R_{giant}}{R_{AdS}}\right)^2=\frac{J}{N}\ ,
\ee%
 where $N$ is number of units of fiveform  flux through the five sphere
in $AdS_5\times S^5$ and \cite{AdS/CFT} \be R^4_{AdS}=R_{S^5}^4=N
l_p^4 \ , \ee where $l_p$ is the ten dimensional Planck length.
The radius of the giant grown inside the $S^5$ cannot exceed its
radius, and hence there is an upper bound on the $J$, $J\leq
N$.\footnote{It is possible to consider giants grown inside
$AdS_5$ \cite{AdS-giants}, for which there is no (upper) limit on
their size and/or angular momentum. However, there is a limit on
the number of such giants \cite{Berenstein, Nemani}, we will
comment more on this issue in section \ref{SYM-gravity-relation}.}

Angular momentum $J$ is quantized and hence there is a minimal
size giant graviton, corresponding to $J=1$ in \eqref{giant-size}.
The size of this object which will be called {\it tiny graviton}
is then given by%
\be\label{tiny-size}%
\left(\frac{R_{tiny}}{R_{AdS}}\right)^2=\frac{1}{N}\ \Rightarrow
R^2_{tiny}=\frac{l^4_p}{R^2_{AdS}}=l^2_p\frac{1}{\sqrt{N}}\ .%
\ee%
Therefore, in the large $N$ (supergravity) limit tiny gravitons
become much smaller than $l_p$. In this limit,%
\be\label{hirarchy}%
R_{giant}\sim R_{AdS} \gg l_p \gg R_{tiny}\equiv l\ . %
\ee%
 We would
like to point out that the sizes for the giant and tiny gravitons
we have discussed above is a classical one. For the giants in
$S^5$ with the radius of order of $R_{AdS}$ the classical
description is a good one (also note that such giants are moving
very slowly, their angular velocity is very small). For the tiny
gravitons, however, the Compton wave-length is much larger than
their classical size. Therefore, the above  arguments should only
be treated as a suggestive one and in fact we need to study the
quantum theory of the tiny gravitons; that is exactly what we are
going to do in the next subsection.

Based on the observation \eqref{hirarchy} it was argued that the
tiny gravitons should then be treated as ``fundamental'' objects
which may be used to formulate a non-perturbative description of
strings on the $AdS_5\times S^5$ or on the plane-wave
\eqref{background}. In other words and in the BFSS terminology,
tiny gravitons are the ``D0-branes'' of the tiny graviton matrix
theory.\footnote{ To complete the above argument, however, one
needs to show that tiny or giant gravitons share another property
with flat D-branes, namely when $J$ number of them become
coincident the $U(1)$ gauge theory living on them enhance to
$U(J)$. Showing this is not as direct as the flat D-branes,
because due to the spherical shape of giants imposing Neumann
boundary conditions only on the directions parallel to the brane
is not as trivial. In \cite{Hedgehog}, using the Born-Infeld
action it was argued that the enhancement of the gauge symmetry
happens. The enhancement of the gauge symmetry for coincident
giants has recently been argued for, using the dual $N=4$ $U(N)$
gauge theory operators corresponding to cioncident giants
\cite{Vijay}.}  The above argument for the three sphere giants can
be repeated for the membrane giants in the $AdS_4\times S^7$,
$AdS_7\times S^4$ or the eleven dimensional plane-wave.
Explicitly,  spherical membranes with a unit angular momentum
become tiny in the large $N$ limit. In \cite{tiny} it was also
noted that the BMN matrix theory \cite{BMN, DSV1} is indeed a
(membrane) tiny graviton theory.

\subsection{Statement of the Conjecture}\label{conjecture-statement}

The tiny graviton matrix theory proposal is  that the DLCQ of
strings on the $AdS_5\times S^5$ or the 10 dimensional \pl \bg\ in
the sector with $J$ units of light-cone momentum, is described by
the theory or dynamics of $J$ ``tiny'' (three-brane) gravitons. To
obtain the action for $J$ tiny gravitons, we follow the logic of
\cite{DSV1} where the corresponding Matrix model is obtained as a
regularized (quantized) version of M2-brane \lc\ Hamiltonian, but
now for 3-branes. This has been carried out in \cite{tiny}. In
other words, DLCQ of type IIB strings on the \pl\ \bg\
\eqref{background} is nothing but a quantized 3-brane theory. The
statement of the conjecture is then:

\begin{quote}
{\it The theory of $J$ tiny three-brane gravitons, which is a
$U(J)$ supersymmetric quantum mechanics with the \super\ symmetry,
is the Matrix theory describing the DLCQ of strings on the
plane-waves or on the $AdS_5\times S^5$ in the sector with
light-cone momentum $p^+=J/R_-$, $R_-$ being the light-like
compactification radius. The \Ham  of this Matrix model is:}
\end{quote}
\be\label{Matrix-model-Ham}
\begin{split}
{\bf H}= R_-\ \Tr&\bl[ \frac{1}{2}\Pi_I^2+
\frac{1}{2}\left(\frac{\mu}{R_-}\right)^2 X_I^2 +\frac{1}{2\cdot
3! g_s^2} [ X^I , X^J , X^K, {\cal L}_5][ X^I , X^J , X^K, {\cal
L}_5]\cr &-\frac{\mu}{3!R_- g_s}\left( \epsilon^{i j k l} X^i
[X^j, X^k, X^l, {\cal L}_5]+ \epsilon^{a b c d} X^a [ X^b, X^c,
X^d , {\cal L}_5] \right)\cr &+\left(\frac{\mu}{R_-}\right)
\left(\psi^\dagger {}^{\alpha \beta} \psi_{\alpha \beta}-
\psi_{\dot\alpha \dot\beta}\psi^\dagger {}^{\dot\alpha
\dot\beta}\right)\cr &+\frac{2}{g_s}\left( \psi^\dagger {}^{\alpha
\beta} (\sigma^{ij})_\alpha^{\:  \: \delta}
  [ X^i, X^j, \psi_{\delta \beta}, {\cal L}_5] +
  \psi^\dagger {}^{\alpha \beta} (\sigma^{ab})_\alpha^{ \: \: \delta} \:
  [ X^a, X^b, \psi_{\delta \beta}, {\cal L}_5]\right) \cr
&-\frac{2}{g_s} \left(\psi_{\dot\delta \dot\beta}
(\sigma^{ij})_{\dot\alpha}^{ \: \: \dot\delta} \:
  [ X^i, X^j, \psi^\dagger {}^{\dot\alpha \dot\beta}, {\cal L}_5]+
\psi_{\dot\delta \dot\beta} (\sigma^{ab})_{\dot\alpha}^{\: \:
\dot\delta} \:
  [ X^a, X^b, \psi^\dagger {}^{\dot\alpha \dot\beta}, {\cal L}_5]\right)\br]\ ,
\end{split}
\ee%
where $I,J,K=1,2\cdots ,8$ and $i,a=1,2,3,4$. In the above
$X^I$'s, $\psi$'s and ${\cal L}_5$ are all $J\by J$ matrices and
the four brackets are the quantized Nambu four-brackets defined
as:%
\be%
[F_1,F_2,F_3,F_4]\equiv \epsilon^{ijkl}F_iF_jF_kF_l%
\ee%
where $F_i$'s are arbitrary $J\by J$ matrices. As discussed in
\cite{tiny} ({\it cf.} Appendix \ref{appendixB}) the quantized
Nambu four brackets are non-associative but satisfy a generalized
Jacobi identity, are traceless and have a by-part integration
property: $Tr[F_1,F_2,F_3,F_4]F_5=-Tr[F_1,F_2,F_3,F_5]F_4 $. Using
the properties of the Nambu four-brackets, one may show explicitly
that the above Hamiltonian exhibits the invariance under the
\super\ superalgebra, which is the superalgebra of the plane-wave
background; see Appendix \ref{superalgebra} for the superalgebra
and its representation in terms of the $J\times J$ matrices. The
other advantage is that, similarly to BMN Matrix model \cite{BMN,
DSV1}, there are no flat directions and the flat directions are
lifted by the mass terms coming form the \bg\ \pl \ metric.

The  $U(J)$ gauge symmetry of the above \Ham\ is in fact a
discretized (quantized) form of the spatial diffeomorphisms of the
3-brane. As is evident from the above construction we expect in
$J\to\infty$ limit to recover the diffeomorphisms. In this
respect, it is very similar to the usual BFSS (or BMN) Matrix
model in which the gauge symmetry is the regularized form of the
diffeomorphisms on the membrane worldvolume \cite{DSV1}.

Here we would like to stress that the DLCQ of strings on the
$AdS_5\times S^5$ and that of the 10 dimensional plane-wave should
be the same. To see this, we note that taking the Penrose limit
over the $AdS_5\times S^5$ we obtain the plane-wave \bg\  and that
the Penrose limit consists of following a light-like observer.
{}From the viewpoint of a light-like observer (a boosted infinite
momentum frame) which uses global AdS time as its time coordinate,
what is seen out of whole $AdS_p\times S^q$ background is the
Penrose limit of that, namely a plane-wave \bg . One should,
however, note that the size of the tiny graviton $l$ which in the
$AdS_5\times S^5$ is given by $l^4=l_p^4/N$ \eqref{tiny-size}, in
the plane-wave limit and in the notations of the Hamiltonian
\eqref{Matrix-model-Ham} is equal to \cite{tiny}%
\be\label{l-fuzziness}%
l^2=\frac{\mu g_s}{R_-}l_s^2.
\ee%
In other words, to use the Hamiltonian \eqref{Matrix-model-Ham}
for the $AdS_5\times S^5$ case one should replace
$\frac{R_-}{\mu}$ by $(g_sN)^{1/2}$. (Note that in our conventions
both $\mu$ and $R_-$ have dimension of energy.)

One less clear ingredient in the TGMT Hamiltonian is a given
classical $J\by J$ matrix ${\cal L}_5$. In sections
\ref{Zero-Energy-section} and \ref{mass-deformed}, we will show
how construction of 1/2 BPS solutions of the TGMT helps us with a
better understanding of the physical meaning of ${\cal L}_5$.

\subsection{Gauge symmetry and the Gauss law constraint}\label{Gauss-section}
The \Ham \eqref{Matrix-model-Ham} can be obtained from a $0+1$
dimensional $U(J)$ gauge theory Lagrangian, in the temporal gauge.
Explicitly, the only component of the gauge field, $A_0$, has been
set to zero. To ensure the $A_0=0$ gauge condition, all of our
physical states must satisfy the Gauss law constraint arising from
equations of motion of $A_0$. Similarly to the BFSS \cite{BFSS}
and BMN \cite{DSV1} cases, these constraints, which consists of
$J^2-1$ independent conditions are: %
\be\label{Gauss-law}%
\left(i[X^i,\Pi^i]+i[X^a, \Pi^a]+
2\psi^{\dagger\alpha\beta}\psi_{\alpha\beta}
+2\psi^{\dagger\dot\alpha\dot\beta}\psi_{\dot\alpha\dot\beta}\right)|\phi\rangle_{phys}=0.
\ee %
These constraints are the requirement of $SU(J)$ invariance of the
physical states.

We should stress that, as in any gauge theory, fixing the local
gauge symmetry does not fix the global gauge symmetry and the
Hamiltonian \eqref{Matrix-model-Ham} is still invariant under the
time independent gauge transformations: \be\label{gauge-trans}
X^I\ \ \rightarrow\ \ U X^I U^{-1}\ \quad ,\quad  {\cal L}_5\ \
\rightarrow\ \ U {\cal L}_5 U^{-1} \ee (and similarly for
fermions) where $U\in U(J)$. That is, ${\cal L}_5$ is a given
matrix up to a $U(J)$ transformation. We will comments on this
issue further in section \ref{Zero-Energy-section}.

\subsection{String theory limit}\label{String-limit}
The \Ham\ \eqref{Matrix-model-Ham} is proposed to describe type
IIB string theory on the \pl with compact $X^-$ direction. The
``string theory limit'' is then a limit where we decompactify
$R_-$, keeping $p^+$ fixed, {\it i.e.}
\be\label{string-theory-limit} J, R_- \to \infty, \qquad \mu,\
p^+=J/R_-, g_s\ \ {\rm fixed}\ . \ee In fact one can show that in
the above string theory limit one can re-scale $X$'s such that
$\mu, p^+$ only appear in the combination $\mu p^+$. Therefore the
only parameters of the continuum theory are $\mu p^+$ and $g_s$.

For the proposal in the full $AdS_5\times S^5$ \bg\ with
$R_-/\mu=\sqrt{g_s N}$ the string theory limit
\eqref{string-theory-limit} is then equivalent to large $N$, large
$J$ ($J,\ N\to \infty$) limit, keeping $J^2/N$ and $g_s$ fixed,
that is the BMN double scaling limit \cite{BMN}.

According to the TGMT conjecture non-perturbative formulation of
type IIB string theory on the $AdS_5\times S^5$ \bg\ in the DLCQ,
is given by quantized D3-brane theory. As a complete theory, TGMT
should also contain other perturbative and non-perturbative
objects present in the string theory, including the fundamental
closed strings themselves. As the second part of the tiny graviton
conjecture it was argued in \cite{tiny} that the ``trivial''
$X^I=0$ vacuum solution of the TGMT, quantum mechanically
describes fundamental IIB closed strings. In other words, here
strings are non-perturbative objects around trivial vacuum.
Providing more supportive evidence for the second part of the
conjecture is postponed to a future work \cite{work-in-progress}.

\section{Zero Energy Solutions}\label{Zero-Energy-section}%
As the kinetic energy is always positive, the zero energy
configurations are necessarily static ($\Pi=0$) solutions,
similarly fermionic terms should also be set to zero. Therefore,
the Hamiltonian relevant for the zero energy solutions takes the
form%
\be\label{vacuum-potential} \begin{split} V = R_-\ \Tr\bl[
&\frac{1}{2}\bl(\frac{\mu}{R_-} X^l + \frac{1}{3!
g_s}\epsilon^{ijkl} [ X^i , X^j , X^k, {\cal L}_5]\br)^2 +
\frac{1}{4g_s^2} [X^i, X^j, X^a, {\cal L}_5][ X^i, X^j,X^a , {\cal
L}_5] \cr + &\frac{1}{2}\bl(\frac{\mu}{R_-} X^d+ \frac{1}{3! g_s}
\epsilon^{abcd}[ X^a , X^b , X^c, {\cal L}_5]\br)^2 +
\frac{1}{4g_s^2} [X^i, X^a, X^b, {\cal L}_5][ X^i, X^a,X^b , {\cal
L}_5] \br]
\end{split}
\ee%
Each term in the above expression is positive-definite, hence the
zero energy solutions are obtained when each of the four terms are
vanishing, {\it i.e.}
\begin{subequations}\label{master-equations}
\begin{align}
 [X^i,X^j,X^k,\L5] & = -\frac{\mu g_s}{R_-}\ \epsilon^{ijkl} X^l \\
[X^a,X^b,X^c,\L5]  & = -\frac{\mu g_s}{R_-}\ \epsilon^{abcd} X^d \\
[X^a,X^b,X^i,\L5]  &= [X^a, X^i, X^j, \L5] =0\ .
\end{align}
\end{subequations}

The first class of solutions to the above equations is the
``trivial'' $X=0$ solution:%
\be\label{trivial} X^i = 0 \quad;\quad X^a = 0
\ee%
Although mathematically trivial, this vacuum is physically quite
non-trivial. According to the tiny graviton conjecture \cite{tiny}
$X=0$ corresponds to the  fundamental string vacuum.

The next class of solutions which was briefly discussed  in
\cite{tiny} is obtained when either  $X^i=0$ or $X^a=0$. In this
case eqs.(\ref{master-equations}c) and either of
(\ref{master-equations}a) or (\ref{master-equations}b) are
trivially satisfied. Since there is a $Z_2$ symmetry in the
exchange of $X^i$ and $X^a$, here we only focus on the $X^a=0$
case and the $X^i=0$ solutions have essentially the same
structure. Therefore, this class of vacua are solutions to
\be\label{single}%
 X^a = 0 \ ,\ [X^i,X^j,X^k,\L5] = -\frac{\mu
g_s}{R_-}\ \epsilon^{ijkl} X^l\ .
\ee%
In sections \ref{single-giant-section} and \ref{concentric} we
give the most general solutions to \eqref{single}. One should,
however, note that if we choose to expand the theory around either
of these vacua the $Z_2$ symmetry is spontaneously broken. As we
will see these solutions are generically of the form of concentric
fuzzy three spheres in either of the $SO(4)$'s.

There is yet another class of solutions where both $X^i$ and $X^a$
are non-zero. These are non-trivial solutions which in the string
theory limit correspond to giant gravitons grown in both $X^a$ and
$X^i$ directions. We consider these cases in sections
\ref{non-concentric} and \ref{generic-multi-giants}.

All of these vacua are 1/2 BPS states. To see this, consider the
superalgebra of the 10-dimensional \pl\bg\ given in Appendix
\ref{superalgebra}. The plane-wave solution \eqref{background} has
a large set of bosonic and fermionic isometries. The bosonic
isometry group, whose dimension is  30, includes  $SO(4)\times
SO(4)$ rotations and translation along $x^-$ and $x^+$ directions,
the generators of which will be denoted by ${\bf J_{ij}}$, ${\bf
J_{ab}}$, $P^+$ and ${\bf H} = -P_-$, respectively \cite{review}.
There are 16 other isometries which are not manifest in the above
coordinates. There are also 32 fermionic isometries (supercharges)
which can be decomposed into 16 kinematical supercharges
$q_{\alpha\beta}$, $q_{\dot\alpha\dot\beta}$ and 16 dynamical
supercharges $Q_{\alpha\dot\beta}$ and $Q_{\dot\alpha\beta}$ (and
their complex conjugates), for more details see \cite{review} and
also Appendix \ref{superalgebra}.

Solutions to eqs.\eqref{master-equations} are {\it fuzzy
3-spheres}, which in the string theory limit generically go over
to giant gravitons (spherical D3-branes) \footnote{ There has been
another proposal for a quantized giant three sphere \cite{Lozano}.
There the fact that $S^3=CP^1\ltimes S^1$ has been used and in the
quantum version the $CP^1=S^2$ has been replaced by its fuzzy
counterpart.}. Noting the equation \eqref{QQ} we see that these
zero energy fuzzy sphere solutions are all $1/2$ BPS, {\it i.e.}
they preserve all of the dynamical supercharges (half out of whole
kinematical and dynamical ones), as they have ${\bf H}=0$ and
${\bf J}_{ij}={\bf J}_{ab}=0 $.

\subsection{Single giant solutions}\label{single-giant-section}
To start our classification of all zero energy 1/2 BPS solutions
of the TGMT,  in this subsection we study a single giant graviton
which is a solution to \eqref{single}. As can already be seen from
\eqref{single} and would be analyzed in detail in the rest of this
section, solutions to \eqref{single} group theoretically are
labeled by $J\by J$ representations of $SO(4)$ (or more precisely
$spin(4)$). These representations can be reducible or irreducible.
The irreducible representations (irreps), which corresponds to a
single giant graviton state, is discussed in this subsection.

Our goal is to solve \eqref{single} for $J\by J$ matrices. Let us,
however, first relax the constraint on the size of the matrices
and look for some generic solution to \eqref{single}. As
\[
[\gamma^i,\gamma^j,\gamma^k, \gamma^5]=-4!\epsilon^{ijkl}
\gamma^l\ ,
\]
it is
straightforward to see that%
 \be\label{4by4-soln}%
 X^i=\zeta
\gamma^i\ ,\qquad \L5=\frac{1}{4!} \gamma^5\ ,%
\ee %
with $\zeta^2=\frac{\mu g_s}{R_-}$ solves \eqref{single}.
($\gamma^i$ and $\gamma^5$ are $4\by 4$ Dirac $\gamma$-matrices.
For our conventions see Appendix \ref{convention}.) Noting that
$\sum_i (\gamma^i)^2=4$, \eqref{4by4-soln} defines a fuzzy three
sphere of radius 2. This is indeed the smallest possible size for
an $S^3_f$.

To construct a generic solution to \eqref{single}, inspired by a
similar method for the fuzzy two sphere \cite{Hammou}, we try to
embed the fuzzy three sphere into a higher dimensional
noncommutative Moyal plane. To do this we introduce an eight
dimensional Moyal plane $\C^4_\theta$, {\it i.e.} a space
parameterized by $z_\alpha,\ {\bar z}_\alpha$, $\alpha=1,2,3,4$
satisfying
\be\label{Moyal}%
[z_\alpha,\ {\bar z}_\beta]=\theta \delta_{\alpha\beta}\ .
\ee%
Next let
\be\label{general-soln}%
X^i=\kappa {\bar z}_\alpha(\gamma^i)_{\alpha \beta}\ z_{\beta}\
,\qquad \L5=\xi {\bar z}_\alpha(\gamma^5)_{\alpha\beta} \
z_{\beta}\ .
\ee%
It can be shown (this has been shown in some detail in Appendix
\ref{appendixB} using identities of Appendix \ref{convention})
that \eqref{general-soln} solves \eqref{single} with
\be\label{normalizations}%
\frac{1}{3} \kappa^2\xi\ \theta^3 (X^0 +2) = \frac{\mu
g_s}{R_-}\equiv {l^2}\ ,
\ee%
where
\be\label{X0}%
 X^0\equiv \frac{1}{\theta} {\bar z}_\alpha{z}_\alpha.
\ee%
The equation \eqref{general-soln} is a realization of the Hopf
fibration for an $S^7$ with an $S^4$ base (for more details see
Appendix \ref{appendixB}).

Noting that the operator $X^0$ commutes with all $X^i$'s and
$\L5$, \eqref{general-soln} is then a solution to \eqref{single}.
In other words, \eqref{general-soln} is a generic solution to
\eqref{single} but with infinite size matrices. Notice that ${\bar
z}_\alpha/\sqrt{\theta}$ and ${z}_\alpha/\sqrt{\theta}$ are
creation and annihilation operators of a four dimensional harmonic
oscillator and hence are explicitly infinite size matrices.
Therefore, $X^i$'s are also infinite size matrices. These infinite
size matrices form {\it reducible} representations of $spin(4)$
and what we should do next is to identify finite size irreps
inside these matrices.

\subsubsection{Restricting to $J\by J$ solutions}\label{reduction-to-JbyJ}%
In this part we construct a finite size solution out of
\eqref{general-soln}. We will do this in two steps; first we
extract out a finite dimensional representation of $SO(5)$ and
then reduce that further to a $J\by J$ representation of $SO(4)$.
Note that $[X^i,X^j]$ are generators of $so(4)$ algebra and adding
$[X^i, \L5]$ to this collection we have a representation of
$so(5)$. Next, note that $X^0$ commutes with all $X^i$'s and $\L5$
and hence with all $so(5)$ generators, {\it i.e.} $X^0$ is a
Casimir of $so(5)$ and $\L5^2$ commutes with all the $so(4)$
generators. Therefore, these two steps can be taken, respectively,
by focusing on $X^0$ and $\L5^2$ and identify the blocks in which
they are proportional to the identity matrix.

In the number operator basis for the four dimensional harmonic
oscillator $X^0$ is already diagonal with the eigenvalues $n$ ($n$
is a non-negative integer). The eigenvalue $n$ comes with the
multiplicity $N$:%
 \be\label{Nvs.n}%
N=\frac{1}{6} (n+1)(n+2)(n+3)\ .%
\ee%
 $N$ is the number of possible partitions of $n$ into {\it
four} non-negative integers. In this basis $X^0$ takes the form
\begin{figure}[h]
\begin{center}
\be\includegraphics[scale=0.55]{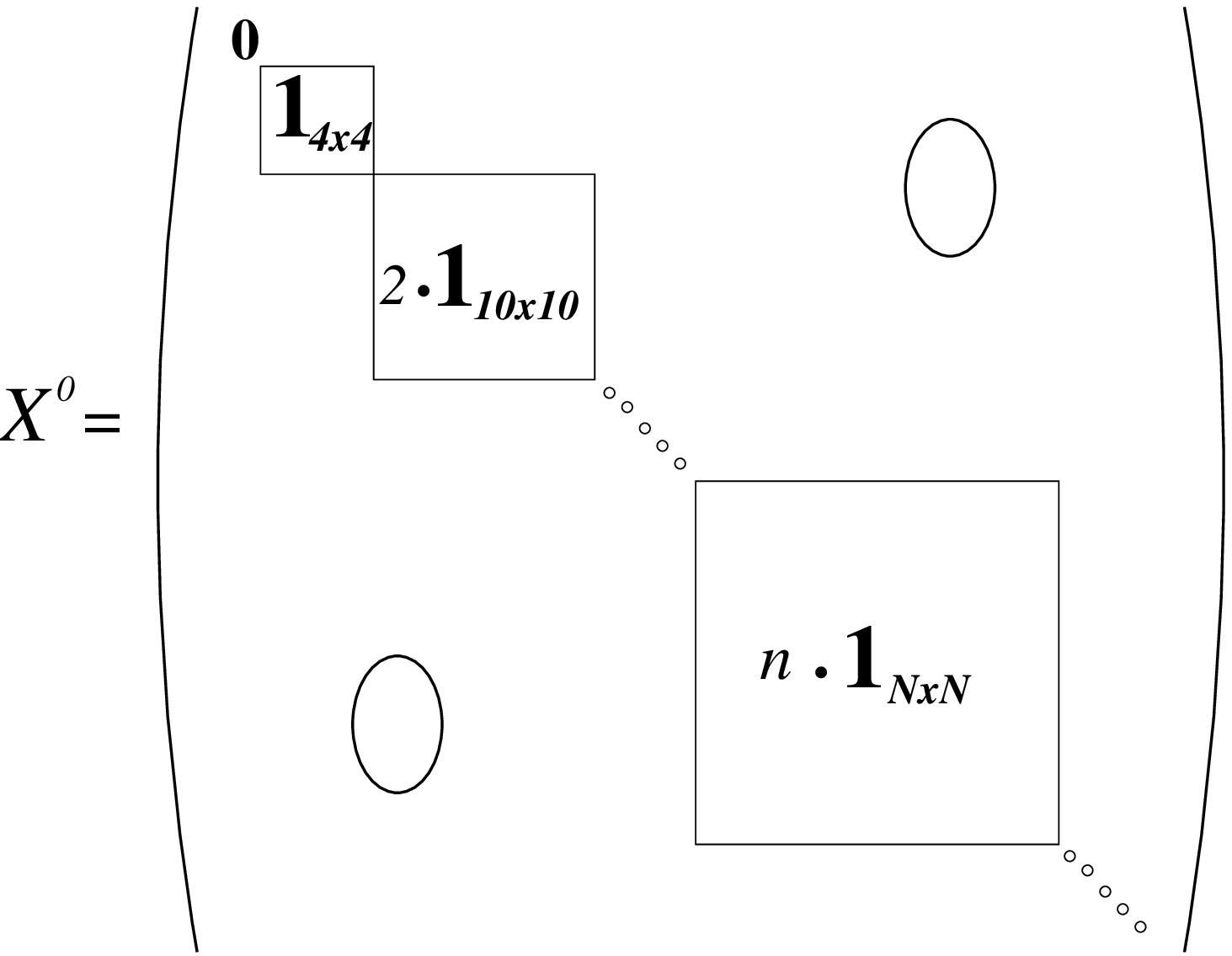}.\ee
\end{center}
\end{figure}

Using the identities of Appendix \ref{convention}, one can easily
check that
\be\label{S4}%
 \sum_{\mu=1}^5 (X^\mu)^2 = \kappa^2\theta^2
X^0(X^0+4)\ , \ee%
where $X^\mu=\kappa {\bar z}\gamma^\mu z$, $\mu=1,2,\cdots, 5$.
Therefore, restricting $X^0$ to a block in which it is equal to
$n\cdot {\bf 1}_{N\by N}$ would give an embedding of a four
sphere, in fact a fuzzy four sphere ({\it cf.} Appendix
\ref{appendixB}), in an eight dimensional Moyal plane. The radius
of this sphere is then
\be\label{S4radius}%
 R^2_{S^4_f}=\kappa^2\theta^2 n(n+4),
 \ee%
  where $n$ and the size of matrices
$N$ are related as in \eqref{Nvs.n}. In the large $n$ limit
$R_{S^4_f}\simeq \kappa \theta\ n $ and $N\simeq n^3/6$. The
continuum (commutative) limit is when $\kappa\theta\sim 1/n \to
0$, keeping $R$ fixed.
\begin{figure}[ht]
\begin{center}
\includegraphics[scale=0.5]{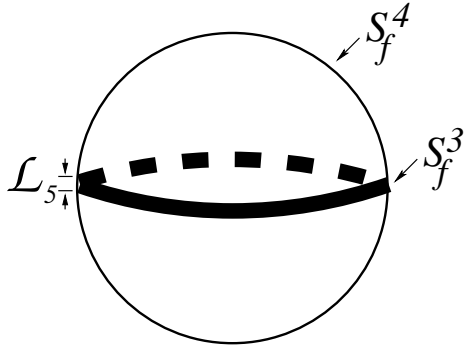}
\caption{A fuzzy three sphere $S^3_f$ is obtained from $S^4_f$ by
cutting it in a narrow strip close to its equator.}\label{S4toS3}
\end{center}
\end{figure}

Before moving to the construction of a fuzzy three sphere, we
would like to comment on the above construction of the four
sphere. By definition an $S^4$ is a four dimensional manifold with
$so(5)$ isometries. In the above we have given a specific
embedding of a four sphere in an eight dimensional
(noncommutative) space. More specifically, noting that $X^0=const.
$ defines an $S^7$ in the eight dimensional space (see
\eqref{X0}), we have an embedding of $S^4$ into $S^7$. This
embedding is a (noncommutative) realization of the Hopf fibration
with $S^4$ as the base, e.g. \cite{Grosse}. Out of the $so(8)$
isometries of the $S^7$ there is a $u(4)$ subgroup which is
compatible with the holomorphic structure on $\C^4\simeq {\mathbb
R}^8$. Note also that in the noncommutative Moyal case of
$\C^4_\theta$, that is this $u(4)\subset so(8)$ which does not
change the noncommutative structure \eqref{Moyal}. The $X^\mu$
behaves as a vector under $so(5)\subset su(4)$ and the generators
of the full $su(4)$ are $X^\mu$ and  $[X^\mu,X^\nu]$. (The
generator of the $u(1)\subset u(4)$ is $X^0$.)\footnote{It is
worth noting that the generators of $so(8)$ which are not included
in $u(4)$ cannot be constructed from combinations of $z,\ \bar
z$.}

We are now ready to take the second step and construct a (fuzzy)
three sphere, $S^3_f$, from the above $S^4_f$. The idea, as
depicted in Figure \ref{S4toS3}, is that a three sphere is a great
sphere on the equator of a four sphere. In the noncommutative
fuzzy case, however, due to the noncommutativity and fuzziness it
is not possible to { exactly} sit on the equator, and we are
forced to cut a narrow strip around the equator. The width of the
strip in the commutative limit goes to zero. This would become
clearer momentarily.

To cut the four sphere around its equator we should restrict the
coordinate $X^5$ to be zero or in fact very close to zero compared
to the other four embedding coordinates. Moreover, by definition
the sum of the squares of the embedding coordinates of an $S^3_f$
must be proportional to the identity matrix. Therefore, we need to
restrict ourselves to blocks in $N\by N$ matrices where $\L5^2$ is
proportional to the identity. In the number operator basis for the
four dimensional harmonic oscillator $X^5$ is diagonal ({\it cf.}
\eqref{general-soln}) and is of the form
\begin{figure}[ht]
\begin{center}
\be\label{X5matrix}
\includegraphics[scale=0.40]{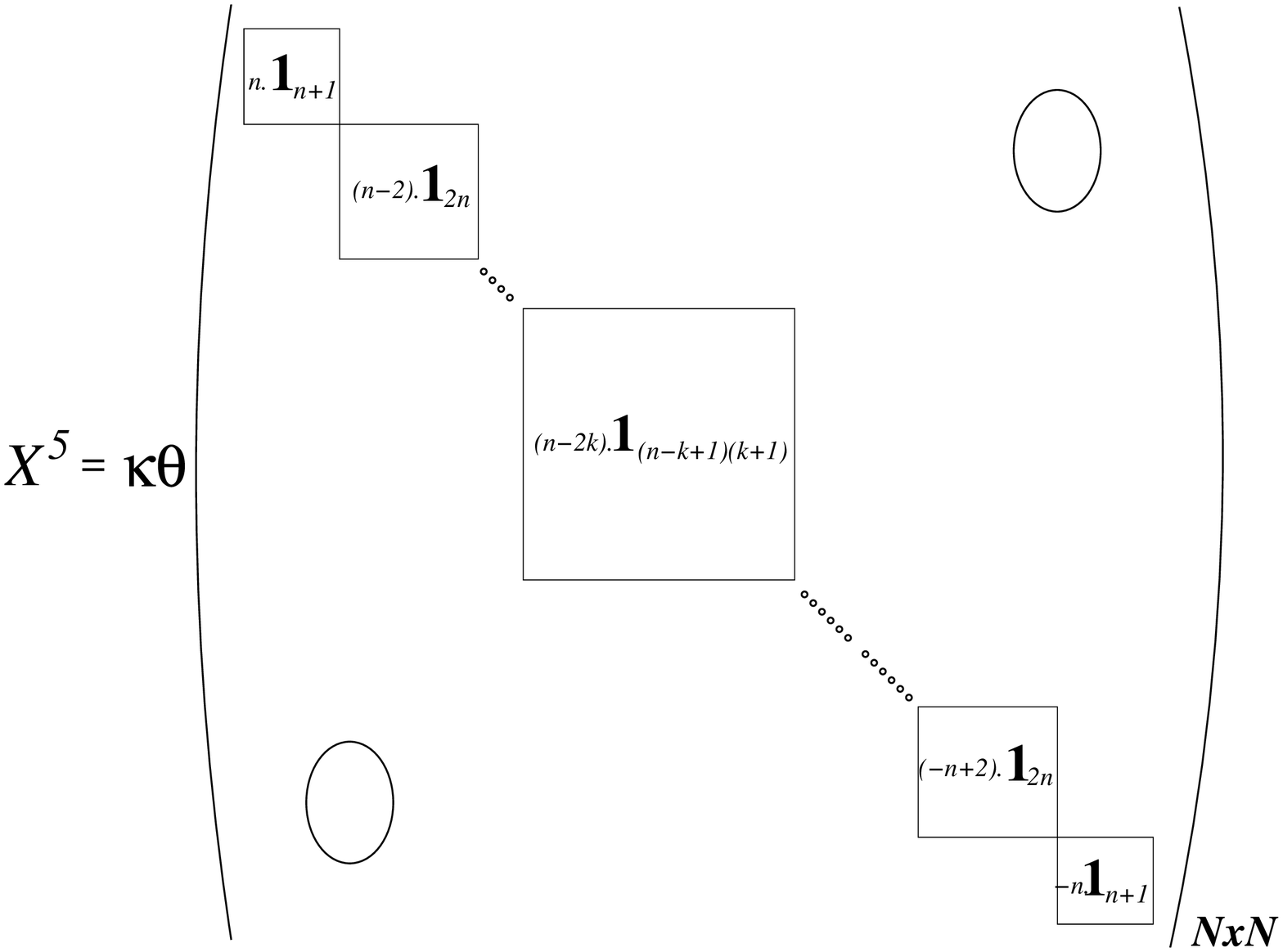},
\ee
\end{center}
\end{figure}
\newline
where $0\leq k\leq n$, {\it i.e.} $X^5$ consists of $n+1$ blocks.

As we see in \eqref{X5matrix} $X^5$ ranges from $-n$ to $n$ with
the steps of two. In the continuum limit this reduces to the fact
that $X^5$ ranges from $-R_{S^4}$ to $R_{S^4}$. The equator then
corresponds to taking the smallest value for $X^5$; that is, zero
if $n$ is even and 1 or $-1$ if $n$ is odd. Obviously even $n$
case, which leads to $\L5=0$, is not an appropriate choice because
all the four brackets in the Hamiltonian \eqref{Matrix-model-Ham}
would vanish. The proper choice is then {\it odd} $n$ case which
leads to the $\L5$ of the form
\begin{figure}[h]
\begin{center}
\be\label{L5matrix}
\includegraphics[scale=0.5]{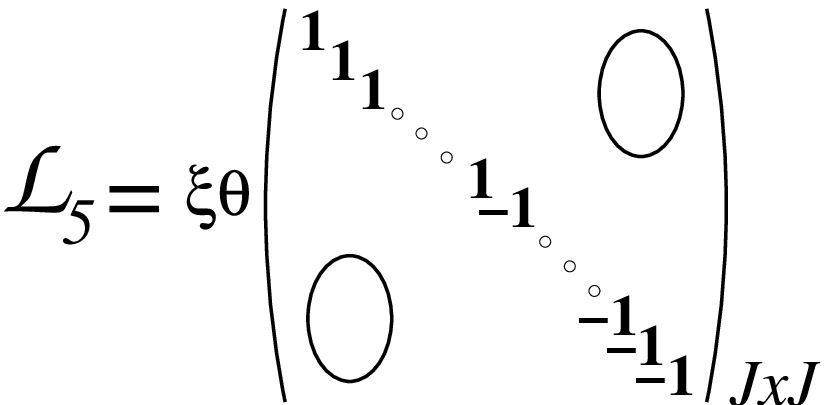}.
\ee
\end{center}
\end{figure}
\newline
The size of this matrix $J$ is
\be\label{Jvs.n}%
J=\frac{1}{2} (n+1)(n+3)\ .
\ee%
This can be seen from the definition of $\L5$ in terms of the
harmonic oscillators and is derived from the number of possible
partitions of a given integer into two non-negative integers.

In \cite{tiny} and thus far, we had not given an explicit basis
independent definition of $\L5$. Using the above, however, now we
have one:

$\L5$ is a $J\by J$ hermitian matrix with
\be\label{L5-def}%
\L5^2= \left({\xi\theta}\right)^2 {\bf 1}_{J\by J}\ \ \ , \qquad  Tr\L5= 0\ .%
\ee%
 Note that
the expression for $\L5$ given in \eqref{L5matrix} is in the basis
where $\L5$ is diagonal and in general $\L5$ is defined up to a
(global) $U(J)$ rotation ({\it cf.} \eqref{gauge-trans}).

To complete our construction of the fuzzy three sphere we should
specify the relation between radius and size of the matrices $J$.
By definition%
\be\label{S3-radius}%
 \sum_{i=1}^4 (X^i)^2 =R^2_{S^3_f} {\bf 1}_{J\by J}\ ,
\ee%
where $X^i$'s are given in \eqref{general-soln}. One may start
with the $N\by N$ matrices corresponding to an $S^4_f$; {\it i.e.}
a spin $n$ representation of $so(5)$ and compute the sum of
squares of $X^i$'s where it becomes proportional to the $J\by J$
identity matrix. In other words, we are projecting $N\by N$
matrices using a projector ${P}_{\cal R}$ where $\sum_{i=1}^4
({P}_{\cal R} X^i {P}_{\cal R})^2 \propto {\bf 1}_{J\by J}$ and
consequently $dim
{P}_{\cal R}= J$. Performing this calculation we find%
\be\label{S3f-radius}%
R^2\equiv R^2_{S^3_f}= \frac{1}{2}(n+1)(n+3)\kappa^2\theta^2= J\kappa^2\theta^2.%
\ee%

In the solution \eqref{general-soln} there are three parameters,
namely $\kappa$, $\xi$ and $\theta$, which are not completely
specified so far. There is, however, a relation among these
parameters and the tiny graviton matrix theory parameters,
\eqref{normalizations} and \eqref{S3f-radius}. Therefore, two of
these parameters, which are normalization parameters, are free and
we have a choice on fixing them. In our conventions $\theta$ has
dimension of length squared and we would like $\L5$ to be
dimensionless and $X^i$ of dimension of length.  Hence, $\kappa$
should have dimension of one over length and $\xi$ dimension of
one over length squared. We choose $\kappa\theta ={l}$, which
leads to $\xi\theta=\frac{l}{R}$. $\theta$ is still a free
parameter which may be absorbed in the redefinition of $z_\alpha$
as $z_{\alpha}/\sqrt{\theta}$. However as we will see in section
\ref{mass-deformed} and is discussed in Appendix \ref{appendixB},
$\theta=l R$ has a natural physical meaning as the minimal volume
which can be measured on a fuzzy four sphere and therefore would
prefer to keep $\theta$ in our formulae. In sum,
we fix our normalization parameters as%
\begin{subequations}\label{normalization-choices}
\begin{align}
\kappa &=\frac{l}{\theta}=\frac{1}{R}\\
\xi & =\frac{l}{R}\cdot \frac{1}{\theta}=\frac{1}{R^2}\\
\theta &= l R\ .
\end{align}
\end{subequations}%
Note that $l^2/l^2_s=\mu g_s/R_-$ \eqref{l-fuzziness}.

With the above choice it is easy to see that typically $X^i\sim R$
while $\L5\sim \frac{l}{R}$. It is now clear that, as we expected
and argued, $\L5$ in the continuum limit, $l\to 0,\ R={\rm
fixed}$, goes to zero. We would like to stress a very important
outcome of the above construction:
\begin{center}
{ A fuzzy three sphere, due to the fuzziness and the fact that
$\L5\neq 0$ (cf. Figure \ref{S4toS3}),\\ is still topologically  a
(fuzzy) four sphere.}
\end{center}
In the continuum limit the width of the $\L5$ strip goes to zero
and we recover the usual three sphere. This would have profound
physical consequences which will be discussed in detail in section
\ref{mass-deformed}.

Before ending this subsection for completeness we briefly review
another equivalent construction of $S^3_f$. A more detailed
discussion on this method can be found in \cite{sunjay1, sunjay2}.
This method is based on group and representation theory of $SO(4)$
and $SO(5)$, rather than the four brackets, embedding into an
eight dimensional Moyal plane and Hopf fibration. For this, let us
start with the construction of the $S^4_f$ presented in Appendix
\ref{appendixB}, \ref{group-theory-construct}. Again, we single
out one of the $\G^\mu$'s, say $\G^5$, and restrict $\G^5$ to a
$J\by J$ sector in which $(\G^5)^2= {\bf 1}_{J\by J}$. The size of
the matrices and $n$, as indicated by the group theory
\cite{sunjay2}, is exactly given by the equation \eqref{Jvs.n}.

\subsection{Multi giant solutions}\label{multi-giant}%
Having studied the single giant graviton solution, we use that to
construct the most general 1/2 BPS solutions, {\it i.e.} multi
giant graviton solutions. This category of solutions to
\eqref{master-equations} encompasses spherical gravitons of
arbitrary size in $X^i$ and/or $X^a$ subspaces. In what follows we
present a detailed study of all possible cases.

\subsubsection{Concentric giants}\label{concentric}%
To go one step further and study more general solutions to what we
have studied so far, in this part we  search for solutions to
\eqref{single} (special case of \eqref{master-equations}) which
geometrically describe a number of concentric spheres in either of
$SO(4)$ invariant subspaces. Here we consider $X^a=0$ and $X^i\neq
0$ case. Group theoretically it corresponds to reducible
representations of $SO(4)$.

\begin{figure}[ht]
\begin{center}
\includegraphics[scale=0.6]{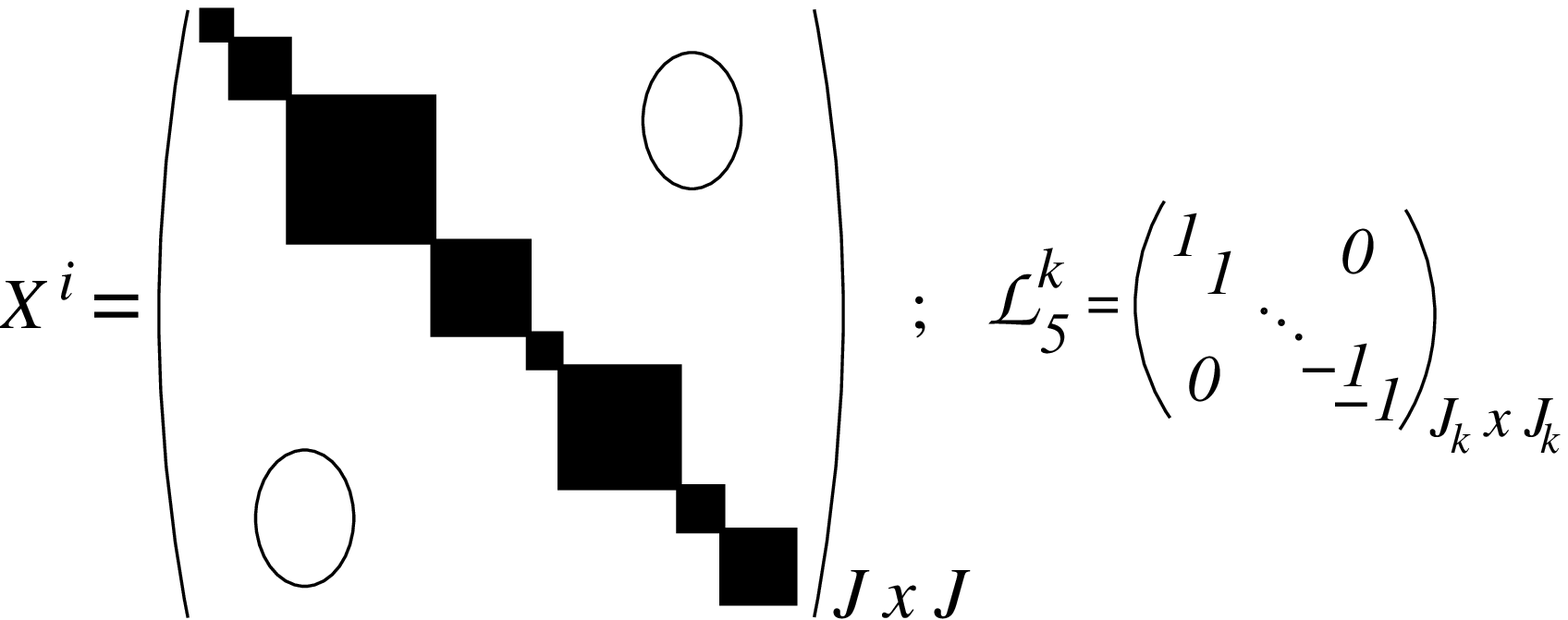}.
\caption{$X^i$ corresponding to $m$ concenteric giants and the
$\L5$ corresponding to the $k^{th}$ block. The size of the
$k^{th}$ three sphere is then $R^2_k\propto J_k$ ({\it cf.}
\eqref{S3f-radius}) and hence $\sum_{k=1}^m R^2_k=
R^2$.}\label{Ximulti}
\end{center}
\end{figure}

This can be obtained by partitioning $J\by J$, $X^i$ matrices into
some $J_k\by J_k$ blocks, $X^i_k$ matrices, in such a way that
$\sum_{k=1}^m J_k = J$,  where $m$ is the number of spherical
gravitons ranging from $1$, corresponding to a single giant giant,
to $J$, corresponding to $J$ tiny gravitons. This has been
depicted in Figure
\ref{Ximulti}.\newline%
\quad We should emphasize that, as solutions to
\eqref{master-equations} and the TGMT, all these solutions must
come with the same $\L5$ matrix. In fact noting the
\eqref{L5matrix} or \eqref{L5-def} one can observe that this is
possible, simply by reshuffling some of $1$'s and $-1$'s in
\eqref{L5matrix}.
\subsubsection{Non-concentric giants}\label{non-concentric}%
As the next  step, we construct solutions to
\eqref{master-equations} where both $X^i$ and $X^a$ are non-zero.
To start with we consider solutions which correspond to one giant
graviton in $X^i$ and one in $X^a$ directions. In order that, we
choose our matrices to have  two copies of the solution we
obtained in section \ref{reduction-to-JbyJ}. To realize this
solution through oscillator approach we have to start from
$\C_\theta^4\times\C_\theta^4$ Moyal complex plane. Since the
construction is essentially the same as what we presented in
\ref{reduction-to-JbyJ}, we do not repeat the details and only
present the final result.

It is easily seen that, by construction, $X^i$ and $X^a$ both
satisfy (\ref{master-equations}a) and (\ref{master-equations}b).
One should, however, make sure that (\ref{master-equations}c) is
also fulfilled. This leads to $X$'s of the form
\begin{figure}[h]
\begin{center}
\be\label{XiXamatrix}
\includegraphics[scale=0.6]{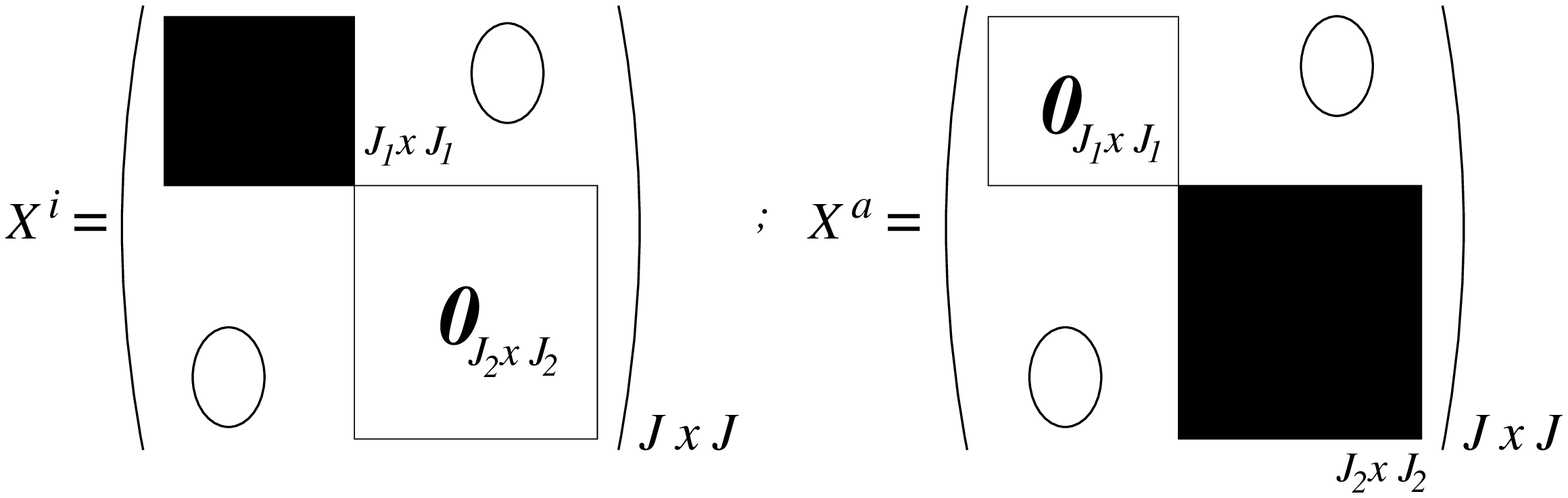}.
\ee
\end{center}
\end{figure}
It is readily seen that $J_1+J_2=J$, or equivalently the sum of
the radii squares of the two spheres should be equal to the square
of the radius of a single giant. It is worth noting that, as it
should, $\L5$ has the same form as in \eqref{L5matrix}, but with
some of the $1$'s and $-1$'s interchanged,
\begin{figure}[h]
\begin{center}
\be\label{L5iL5amatrix}
\includegraphics[scale=0.75]{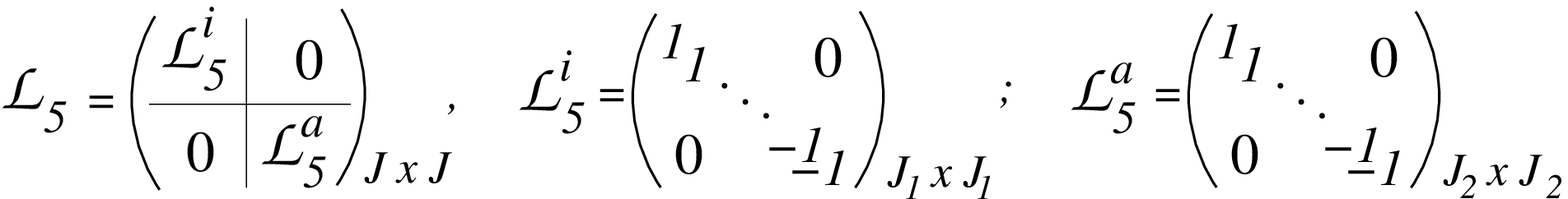}.
\ee
\end{center}
\end{figure}
\subsubsection{Generic multi giants}\label{generic-multi-giants}%
\begin{figure}[ht]
\begin{center}
\includegraphics[scale=0.5]{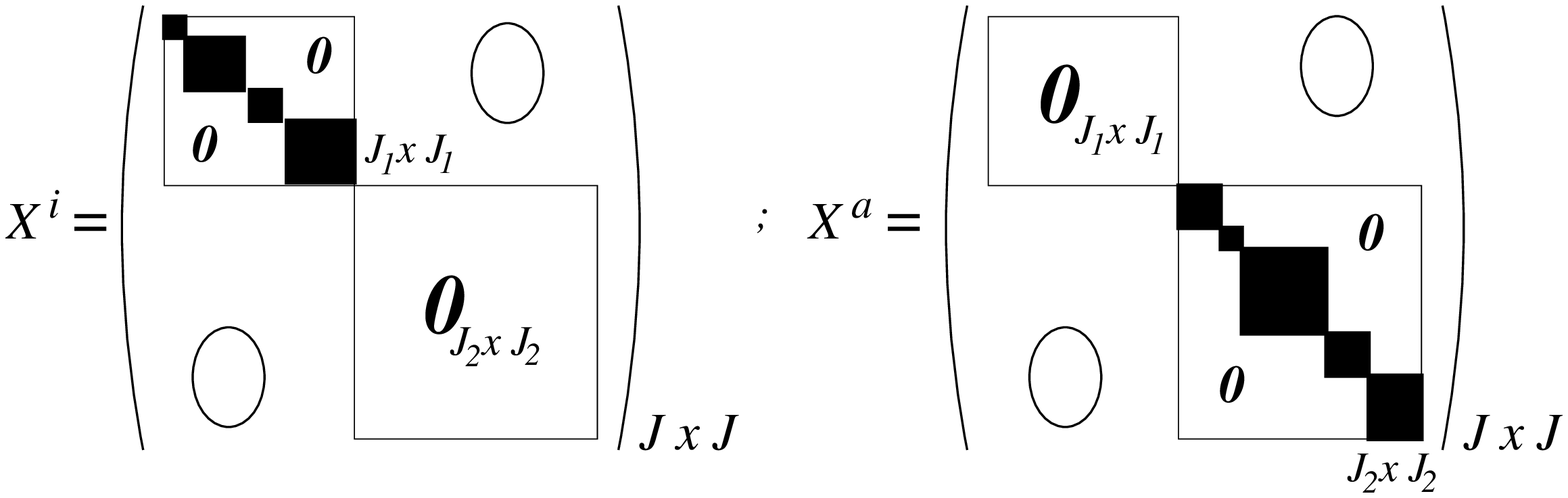}.
\caption{$X$'s corresponding to the most generic 1/2 BPS solutions
of TGMT.} \label{XiXamultimatrix}
\end{center}
\end{figure}

Finally we can combine the arguments of \ref{concentric} and
\ref{non-concentric} sections to construct the most generic
solutions to \eqref{master-equations}. These solutions describe
arbitrary number of $S^3$ giants in both $X^i$ and $X^a$
directions. As depicted in Figure \ref{XiXamultimatrix},
coordinates of each spherical graviton come from each block of the
partitioned $X^i$ and $X^a$ matrices and the radius squared of
each $S^3$ is proportional to  the size of the corresponding
block.

It is instructive to compare our solutions to that of BMN matrix
theory \cite{BMN} which, as discussed in \cite{tiny}, is nothing
but (membrane) tiny graviton matrix theory. There we have  $X^i$
and $X^a$ directions but with $i=1,2,3$ and $a=1,2,\cdots ,6$. The
zero energy solutions are {\it either} of the form of concentric
fuzzy two sphere membranes grown in the $X^i$ directions
\cite{DSV1} {\it or} concentric spherical fivebranes grown in the
$X^a$ directions \cite{MSV}. In other words, each vacuum has a
description in terms of membranes or equivalently described by
fivebranes. In \cite{MSV} it was argued that both the membrane and
fivebrane vacua can be described by a Young tableau encoding
partition of the light-cone momentum $J$ into non-negative
integers. Depending whether we focus on the rows or columns of the
Young diagram we will see the membrane or the fivebrane
description. In this respect this is very similar to our case
({\it cf.} Figure \ref{AdSvsS-giants}), however, now both the
$X^i$ or $X^a$ directions correspond to spherical three branes.

Compared with the 11 dimensional case of the BMN matrix theory
\cite{DSV1, MSV}, however, the ten dimensional case of this paper
has some specific features.  In the ten dimensional case, as we
have shown one can explicitly construct solutions in which  both
$X^i$ and $X^a$ are non-vanishing. (In fact to the authors'
knowledge for the 11 dimensional case so far there is no explicit
construction of the fivebranes in terms of matrices.) Moreover, in
the ten dimensional case, besides the two possible dual
descriptions in terms of the three sphere giants grown in $X^i$ or
$X^a$ directions, there should be yet another description in terms
of fundamental strings \cite{work-in-progress}.

\section{Relation to 1/2 BPS States of $\N=4$ $U(N)$ SYM and Their
Gravity Duals}\label{SYM-gravity-relation} The tiny graviton
matrix theory is conjectured to describe DLCQ of strings on the
$AdS_5\times S^5$ background. The latter, however, has another
description in terms of $\N=4$, $D=4$ $U(N)$ SYM theory. As such,
we should be able to show that there is a one-to-one map between
the 1/2 BPS solutions of the two. For that, let us first review
the structure of 1/2 BPS solutions in the $U(N)$ SYM.  The $\N=4$,
$D=4$ $U(N)$ SYM is a superconformal field theory and has a large
supersymmetry group, namely $PSU(2,2|4)$, the  bosonic part of
which is $SO(6)\times SO(4,2)$. All the (gauge invariant)
operators of the gauge theory fall into various (unitary)
representations of this supergroup.  Starting from the
$psu(2,2|4)$ in the appropriate notation, e.g. the one adopted in
\cite{review}, it is straightforward to show that 1/2 BPS states
are those which are invariant under $SO(4)_i\subset SO(4,2)$ and
$SO(4)_a\subset SO(6)$. Besides this $SO(4)\times SO(4)$,  1/2 BPS
operator must be invariant under another $U(1)$, $U(1)_H$. This
$U(1)$ is constructed from $U(1)_{\Delta}$ and $U(1)_J$, where
$SO(4)_i\times U(1)_\Delta\subset SO(4,2)$ and $SO(4)_a\times
U(1)_J\subset SO(6)$, as follows: denote the generators of
$U(1)_\Delta$ and $U(1)_J$ by $\Delta$ and $J$, the generator of
$U(1)_H$ is then $H=\Delta-J$. (Note that $U(1)_\Delta$ is a
non-compact $U(1)$ while $U(1)_J$ is a compact one.) Therefore,
1/2 BPS states should have $H=0$ and be invariant under
$SO(4)\times SO(4)$. In other words, 1/2 BPS states of the $\N=4,\
D=4$ SYM from the superalgebra viewpoint have exactly the same
quantum numbers as the zero energy solutions of the tiny graviton
matrix theory and naturally fall into the unitary representations
of \super .

\begin{figure}[ht]
\begin{center}
\includegraphics[scale=0.9]{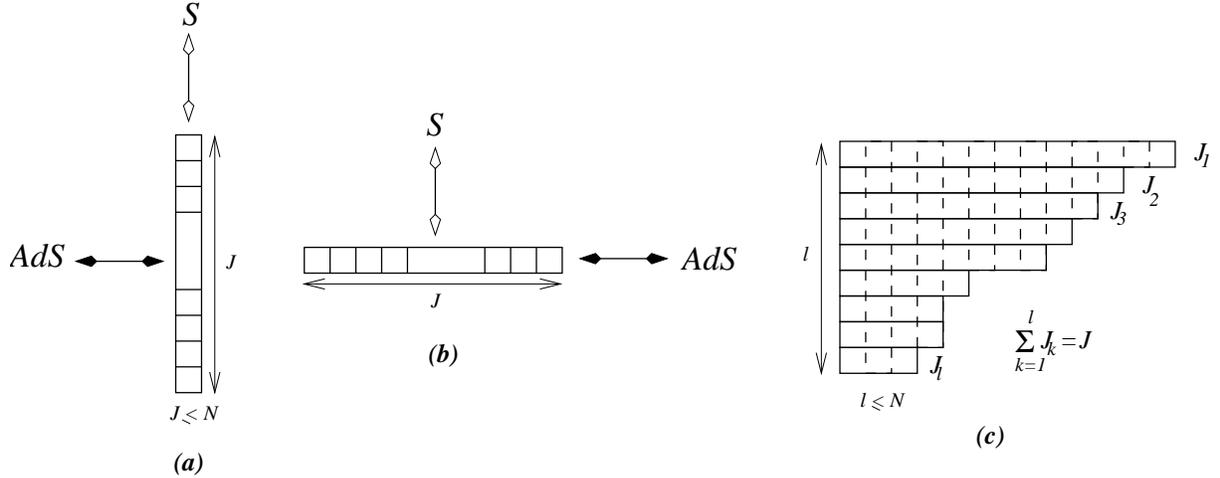}
\caption{Some Young Tableaux with $J$ boxes. (a) A totally
anti-symmetric representation. This corresponds to a single giant
graviton of radius $\sqrt{J}$ grown in $S^5$ {\it or} $J$ tiny
gravitons residing in $AdS_5$. (b) A totally symmetric
representation which corresponds to a single giant graviton in
$AdS_5$ {\it or} $J$ tiny gravitons in $S^5$. (c) A generic Young
tableau of $l$ rows and $J$ boxes. If we view the tableaux from
above (focusing on columns) we see giants grown in $S^5$, and if
we view it from the left side (focusing on rows) we see giants in
$AdS_5$.  In a $U(N)$ Young tableau number of rows cannot exceed
$N$, a realization of the stringy exclusion principle
\cite{MST}.{} From the viewpoint of giants grown in $AdS$,
however, this is the number of concentric 3-branes, and not their
size, which cannot exceed $N$. The fact that each Young tableau
has two interpretations in terms of giants in $S^5$ or $AdS_5$ is
a manifestation of particle-hole duality in the two dimensional
fermion picture discussed in \cite{LLM}.} \label{AdSvsS-giants}
\end{center}
\end{figure}

After this strong indication coming from the superalgebra and its
representations, let us build a more direct relation between our
fuzzy sphere solutions and the gauge theory 1/2 BPS operators. In
the $\N=4$ gauge multiplet we have six real scalars $\phi_A$ which
form a vector of $SO(6)$. Let $\phi_5+i\phi_6=Z$. A generic 1/2
BPS operator, a chiral primary operator, is then a multi-trace
operator only made out of $Z$'s: \be\label{chiral-priamry} {\cal
O}_{\{k_i\}}=\ :TrZ^{k_1}\  TrZ^{k_2}\cdots  TrZ^{k_l}\ : \ee
where the total R-charge (or scaling dimension) of ${\cal
O}_{\{k_i\}}$ is $J=\sum_{i=1}^l k_i$. (There might be some
repeated $k_i$'s.) The ``Tr'' basis is not necessarily an
appropriate one, ${\cal O}_{\{k_i\}}$ and ${\cal O}_{\{k'_i\}}$
are not orthogonal to each other. Instead one may use
``subdeterminant'' basis \cite{Vijay, Bala-Strassler} or ``Schur
polynomial'' basis \cite{Jevicki, holoshape, deMelloKoch} which
have the orthogonality condition. Consider a Young tableau
consisting of $J$  boxes, this Young diagram represents both a
representation ${\cal R}$ of $U(N)$, and also a representation of
the permutation group of $J$ objects ${\cal S}_J$, $\chi_{\cal
R}$. Hence one may use this observation to construct the Schur
polynomial basis:%
\be\label{Schur-basis}%
 {\cal O}_{{\cal
R}}=\frac{1}{n!}\ \sum_{{i_1}, i_2,\cdots , i_J}\ \
\sum_{\sigma\in {\cal S}_J} \chi_{{\cal R}} \ \
\left(Z^{i_1}_{i_{\sigma(1)}}Z^{i_2}_{i_{\sigma(2)}}\cdots
Z^{i_J}_{i_{\sigma(J)}}\right)\ .%
\ee %
A generic Young diagram is depicted in Figure \ref{AdSvsS-giants}.
In other words, there is a one-to-one map between representations
of permutation group ${\cal S}_J$ and the chiral primary
operators.

On the other hand there is a one-to-one correspondence between
partitions of $J$ into arbitrary positive integers and the
representations of ${\cal S}_J$. Hence it is evident that there is
a one-to-one correspondence between our generic fuzzy sphere
solutions of section \ref{generic-multi-giants} and the Young
Tableaux with $J$ boxes. In this point of view each box in the
Young tableau corresponds to a tiny graviton.

\begin{figure}[ht]
\begin{center}
\includegraphics[scale=0.95]{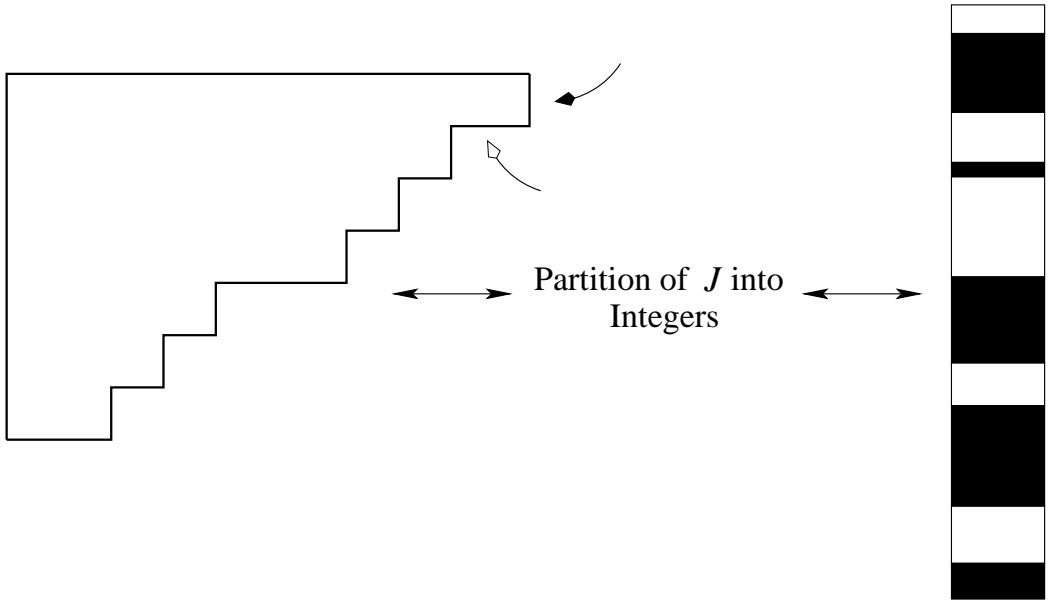}
\caption{A Young tableau of $J$ boxes has three interpretations.
{\it i}) A representation of $U(N)$, {\it ii}) Schur Polynomials
and representation of permutation group ${\cal S}_J$ and {\it
iii}) Partition of $J$ into non-negative integers. The first two
interpretations were noticed in \cite{LLM, Berenstein,
deMelloKoch} as discussed there, they lead to an equivalent
description of all 1/2 BPS configurations of $\N=4$ $U(N)$ SYM in
terms of two dimensional fermion system. The third one, however,
is the one, is relevant to the 1/2 BPS
solutions of the TGMT.} \label{tableau-partion}
\end{center}
\end{figure}
We would also like to briefly comment on the relation to the
supergravity solutions corresponding to the above 1/2 BPS states,
recently constructed by Lin-Lunin-Maldacena \cite{LLM}. However,
first we need to review another equivalent description of the 1/2
BPS opterators of the $\N=4$ $U(N)$ gauge theory: A system of $N$
two dimensional fermions in harmonic oscillator potential. To see
this recall that the part of the $\N=4$ $U(N)$ gauge theory on
$R\times S^3$, relevant for the dynamics of  chiral primary
operators, in the temporal gauge, takes a very simple form:%
\be\label{Z-action}%
 S=\frac{1}{g^2_{YM}} \int d\tau \
Tr\left(\partial_\tau Z^\dagger\partial_\tau Z- Z^\dagger
Z\right)\ .
\ee%
 The simplifications leading to the above action
come from two sources: 1) the chiral primaries, among all the
fields present in a $\N=4$ gauge multiplet, only involve $Z$'s
and; 2) to be 1/2 BPS they should be invariant under $SO(4)$
acting on $S^3$ that the gauge theory is defined on. Therefore, we
can perform integration over the $S^3$ to remain with a one
dimensional  action \eqref{Z-action}. The mass term for $Z$ is the
conformal mass present for all of six scalars (and fermions) of
the $\N=4$ SYM on $R\times S^3$. The action \eqref{Z-action} which
governs the dynamics of the chiral primary operators is nothing
but the action for a bunch of decoupled harmonic oscillators all
with the same frequency (note that $Z$'s are $N\by N$ matrices).
One can still use the (global) $U(N)$ gauge transformations to
bring $Z$'s to a diagonal form. These $N$ eigenvalues can directly
be related to physical gauge invariant operators. In fact the
eigenvalues correspond to position of $N$ two dimensional fermions
in a harmonic oscillator potential. (If we choose to work with
diagonal matrices, we should then add the Van der monde
determinant, which in turn leads to the fermionic nature of the
eigenvalues \cite{Brezin}.)

In the two dimensional fermions viewpoint every chiral primary
corresponds to a distribution of fermions or a ``droplet'' in the
fermion phase space. In a recent work Lin-Lunin-Maldacena (LLM)
\cite{LLM} have constructed the supergravity solutions
corresponding to each droplet. That is, they have shown that there
is a one-to-one map between the 1/2 BPS type IIB supergravity
solutions with \super\ supersymmetry and the $N$ 2d fermion phase
space. These solutions, as supergravity solutions, correspond to
various {\it classical} giant gravitons in the $AdS_5\times S^5$
or the plane-wave \bg . We have discussed that there is a
one-to-one correspondence between  our 1/2 BPS solutions and the
$\N=4$ gauge theory, hence we expect that our multi-fuzzy sphere
solutions to be associated with the {\it quantized}  LLM
backgrounds. It is, however, hard to directly construct classical
\bg\ metrics corresponding to a given Matrix theory
configurations. This is also the case for BFSS Matrix Theory or
its variants. Nonetheless, it should be possible to extract some
information about the blackholes and their entropy from the matrix
theory. We hope to address this question in future works.

\section{Relation to Mass Deformed $D=3,\ \N=8$ SCFT}\label{mass-deformed}
As discussed in \cite{tiny}  and reviewed in section
\ref{review-section}, the TGMT Hamiltonian is obtained  from
quantization (discretization) of a {3-brane} Hamiltonian in the
plane-wave background, once the light-cone gauge is fixed. In the
process of quantization, we prescribed to replace the classical
Nambu 3-brackets, which naturally appear in the 3-brane action,
with quantized Nambu 4-brackets. For that, however, we need to
introduce a given fixed matrix, $\L5$. In this section we intend
to clarify the role and physical significance of the $\L5$ and
justify our prescription for quantization of Nambu 3-brackets.

In order that we recall the standard string/M-theory dualities and
that type IIB string theory is related to M-theory on a $T^2$,
under which an M5-brane (wrapping the $T^2$) is mapped to a
D3-brane. On the other hand, the TGMT is a quantized 3-brane
theory. Therefore, we propose that:
\begin{quote}
{ The TGMT can also be thought as a quantized M5-brane theory,
$\L5$ is indeed the reminiscent of the 11$^{th}$ circle and the
three sphere giants are M5-branes with the worldvolume ${\mathbb
R}^{2+1}\times S^3$. The ${\mathbb R}^2$ part is in fact
compactified over the $T^2$ which relates M and type IIB theories,
one of the directions in  $T^2$ is then along the $\L5$ and the
other one is along the $X^-$ direction.}
\end{quote}

\subsection*{\ \ \ \ Evidence for the proposal}
To present some pieces of evidence in support of the above
proposal we first, very briefly, review the results and arguments
of Bena-Warner \cite{BW}, which may also be found in \cite{LLM},
where a class of 1/2 BPS solutions to 11 dimensional supergravity
has been constructed. These solutions asymptote to $AdS_4\times
S^7$ and, as discussed in \cite{BW}, correspond to M2-branes
dielectrically polarized into M5-branes. In these solutions the
$S^7$ has been deformed in such a way that its $SO(8)$ isometry is
reduced to $SO(4)\times SO(4)$. Physically these solutions
correspond to (near horizon geometry of) a stack of M2-branes
along $X^0,X^1, X^2$ directions and M5-branes along $X^0,X^1,X^2$
while their other three directions wrap an $S^3$ in the directions
transverse to the M2-branes. Hence, the worldvolume of the
M5-branes is on ${\mathbb R}^{2+1}\times S^3$. {}From the gravity
viewpoint this deformed $AdS_4\times S^7$ is obtained by turning
on a four-form flux through either of the $S^3$'s and the other
remaining direction in $S^7$. (In our analysis we take this
${\mathbb R}^2$ to be compactified on an $S^1\times S^1$.) Due to
the existence of the four-form flux one of this $S^1$ directions
combine with one of the $S^3$'s to {topologically} form an $S^4$;
the M5-branes then wrap the other $S^3$ \cite{LLM}.

This deformation of $AdS_4\times S^7$ has a dual description in
terms of mass deformed three dimensional ${\cal N}=8$ SCFT. The
mass deformation, which corresponds to the four-form flux through
the $S^3$ in the supergravity solution, breaks the $SO(8)$
$R$-symmetry of the SCFT to a $SO(4)\times SO(4)$ subgroup, as
well as half of the supersymmetry.

The isometries of this solution include two translations along
the M2-brane worldvolume. Following \cite{BW, LLM}, one may
U-dualize this solution along this ${\mathbb R}^2$ down to type
IIB supergravity. In the type IIB, however, the M5-branes appear
as three sphere giant gravitons. One of the U-duality directions
after the duality appear as the $X^-$ direction in our ten
dimensional plane-wave \bg . In our DLCQ description this
direction is compact with the radius $R_-/\mu$ (in units of
$l_s$). We propose that the other one has a non-trivial appearance
in our TGMT Hamiltonian through the $\L5$. In other words, the
TGMT already knows about the 11 dimensional origin of the type IIB
string theory.

As the first piece of evidence recall that, as we discussed and
emphasized in section \ref{single-giant-section}, a fuzzy three
sphere which is a {\it quantized} three sphere giant, is
topologically a four sphere. The $\L5$, which is the reminiscent
of the (fuzzy) four sphere we started with, has a non-vanishing
extent, {\it cf.} Figure \ref{S4toS3}. The non-zero extent of
$\L5$ is a ``quantum'' effect and therefore the 11 dimensional
origin of the TGMT is a result of {\it quantization} of type IIB
string theory. Recalling that $\L5\sim l/R$, in the classical
limit, $l\to 0$, $R$=fixed, $\L5$ goes to zero and hence we
reproduce the usual ten dimensional type IIB theory.

As a more quantitative evidence, we note the relation between the
fuzziness, $g_s$ and $R_-$ \eqref{l-fuzziness}, which can be
written as \be\label{gs-Rmul}
g_s=\frac{{R_-}/{\mu}}{{l^2}/{l_s^2}}. \ee The above should be
compared with the expression for the coupling of type IIB string
theory obtained from M-theory on $T^2$. If the radii of the two
cycles in $T^2$ are $R_1, R_2$, then $g_s=R_1/R_2$, e.g. see
\cite{Townsend}. If we identify $R_1$ with $R_-/\mu$, then
\eqref{gs-Rmul} implies that $R_2=l^2/l_s^2$. \footnote{Here we
assume that $R_1$ and $R_2$ are both measured in the same units,
e.g. $l_s$ or 11 dimensional Planck length and hence work with
dimensionless quantities.}

Next, we note that as a direction in a fuzzy four sphere the width
of the $\L5$ strip in units of $l_s$ is $\delta X^5=L\L5\sim L
l/R$, where $L^4$ is the smallest volume one can measure on an
$S^4_f$, {\it cf.} discussions at the end of Appendix
\ref{appendixB}. Using
\eqref{L4=l2R2} we have%
\be\label{deltaX5}%
 \delta X^5=\frac{lR}{l_s^2}\cdot \frac{l}{R}=\frac{l^2}{l_s^2}\ ,%
 \ee%
which is what we expected for $R_2$ from a $T^2$ compactification
of M-theory. This is a remarkable result because reproduction of
the standard string/M-theory dualities and their implications is
among the strong supportive evidence for the TGMT conjecture.

\section{Discussion and Outlook}\label{discussion-section}

As a start to a detailed analysis of the tiny graviton matrix
theory (TGMT), in this paper we studied the zero energy 1/2 BPS
configurations of the theory. These solutions are labeled by
reducible or irreducible $J\by J$ representations of $SO(4)$ and
physically correspond to various giant gravitons of different
size, extended in $X^i$ and/or $X^a$ directions.

\begin{figure}[ht]
\begin{center}
\includegraphics[scale=1.1]{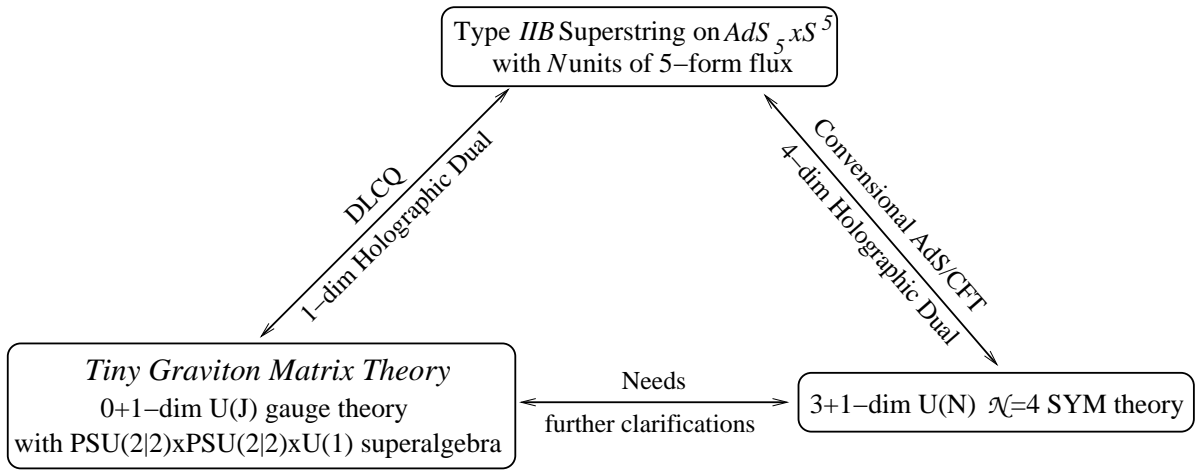}
\caption{The AdS-TGMT-CFT triality}\label{triality}
\end{center}
\end{figure}

Recently there have been an interest in better understanding of
similar 1/2 BPS configurations  in the context of usual AdS/CFT
correspondence in both gauge theory \cite{Berenstein} and gravity
\cite{LLM} sides. The interest in the 1/2 BPS configurations is
motivated by the fact that these 1/2 BPS states may help us with
pushing the analysis of AdS/CFT to beyond the supergravity limit.
In this respect our analysis parallels these studies, though in
the context of the TGMT which is conjectured to be a
non-perturbative, exact description of DLCQ of type IIB strings on
$AdS_5\times S^5$. Of course, analysis of the 1/2 BPS states is a
part of a bigger problem, which has been depicted in Figure
\ref{triality}. In \cite{tiny} and also in this work we have tried
to, at least, start addressing the AdS-TGMT-CFT triality. In this
direction we have noted that a Young tableau can simultaneously
label all the 1/2 BPS states/configurations of the $\N=4, D=4$
$U(N)$ SYM, type IIB supergravity \cite{LLM} and also TGMT, see
Figure \ref{tableau-partion}. It is very desirable to expand our
understanding of the triality, which is postponed to future works.
One interesting question is whether the two dimensional fermions
of \cite{LLM, Berenstein} has a closer relation to the tiny
gravitons \cite{work-in-progress}. Among the points and puzzles to
be understood in this direction we should mention the fact that
size of the matrices corresponding to a single fuzzy sphere
solution, $J$, is fixed once we identify a quantum giant graviton
as a fuzzy three sphere. In particular $J$ cannot take any
arbitrary value ({\it cf.} eq.(3.16)). This rises the puzzle that
in the gauge theory we can have e.g. subdeterminant operators with
arbitrary length (or R-charge). Resolution of this puzzle is
postponed to a future work \cite{work-in-progress}.

\begin{figure}[ht]
\begin{center}
\includegraphics[scale=1]{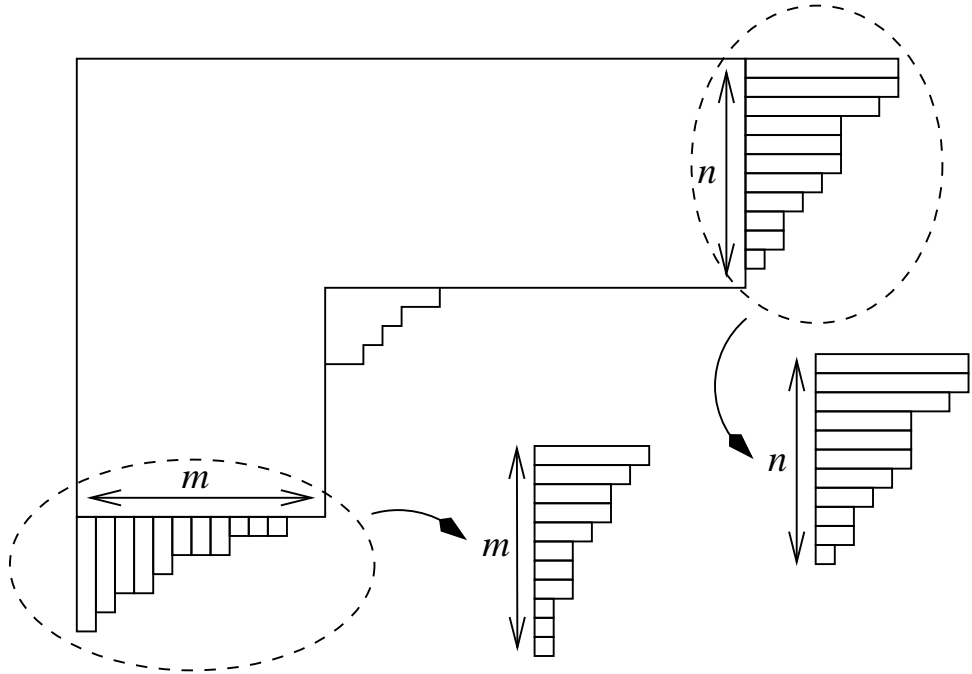}
\caption{The ``fluctuations'' above the Young tableau
corresponding to $n$ coincident giants in $AdS$ or $m$ coincident
giants in $S^5$ can be thought as a Young tableau for states in a
$U(n)$ or $U(m)$ gauge theory.}\label{Young-tableau-U(m)}
\end{center}
\end{figure}

One of the basic ingredients used in the formulation of  TGMT is
the fact that the gauge symmetry of $m$ coincident giant
gravitons, similarly to the flat D-branes, is $U(m)$. There have been
two independent arguments in support of  this property
\cite{Hedgehog, Vijay}. Noting a generic Young tableau we already
see strong indications for this, see Figure
\ref{Young-tableau-U(m)}. It is, however, instructive to re-derive
the same fact from the TGMT. As the first step in this direction
one may start with working out a configuration in TGMT
corresponding to the giant hedge-hogs of \cite{Hedgehog}, the
``fuzzy giant hedge-hogs''.

As we have seen so far giant gravitons, and not fundamental
strings of type IIB, naturally appear in the TGMT. The first
question to ask is where are the type IIB fundamental strings. In
\cite{tiny} there was a proposal that the $X=0$ vacuum corresponds
to single string vacuum, around which fundamental strings appear
as non-perturbative objects. In addition to the single string
vacuum, we need to work out the details of multi-string vacua.
Moreover, as TGMT is a formulation of type IIB string theory with
compact $X^-$ direction, to specify the string vacuum states
besides the light-cone momentum we should also give the winding
number.\footnote{ We would like to thank Juan Maldacena for his
comment on this issue.} Besides the vacuum states it is
interesting to make connection between the TGMT description of
type IIB strings on the ten dimensional plane-wave and the string
bit model of Verlinde \cite{Stringbits}.

In section \ref{review-section} we gave an argument that the {\it
classical} size of a tiny graviton is $l$. Through a detailed
analysis of the fuzzy three sphere solutions, however, we found
that {\it quantum} mechanically the smallest volume that one can
probe on an $S^3_f$ is in principle much larger than $l^4$, that
is $L^4$ \eqref{L4=l2R2}. It is interesting to compare $L$ with
the ten dimensional Planck length $l_p$. Combining,
\eqref{tiny-size},
\eqref{L4=l2R2} and \eqref{S3f-radius} we obtain%
\be\label{Lvs.lp}%
L^2=\sqrt{\frac{J}{N}}l^2_p\ . %
\ee%
When the stringy exclusion relation discussed in \cite{MST} is
saturated, {\it i.e.} $J\sim N$, $L\sim l_p$ and hence in this
case $l_p$ is indeed the smallest physical length that one can
probe.

In constructing our solutions we used an embedding of the fuzzy
three spheres in  $\C^4_\theta\times \C^4_\theta$. It is amusing
to see whether these two copies of the eight dimensional Moyal plane are
just mathematical tools or have a more physical meaning. As a
relevant point we note that the minimal physical length scale of
the TGMT is $L$ which is equal to the noncommutativity
scale on these Moyal planes \eqref{theta-L2}.

In the Appendix \ref{harmonic-oscil-construct} we reviewed and
elaborated on a new method of construction of fuzzy spheres based
on embedding them into higher dimensional Moyal planes. This
method works perfectly well for the cases where the sphere admits
a Hopf fibration. There are, however, only four such possibilities
somehow related to the fact that the division algebras are only
limited to real ${\mathbb R}$, complex $\C$, quaternion ${\mathbb
H}$ and octonions ${\mathbb O}$. Among these cases we have used
the first three and only the last one, which leads to a Hopf
fibration of an $S^{15}$ with an $S^8$ base is remaining. One may
use this and discussions of \ref{harmonic-oscil-construct} to
construct a description of $S^8_f$ and $S^7_f$ by embedding them
into a $\C^8_\theta$.

As discussed in section \ref{mass-deformed} the TGMT is related to
a mass deformed $D=3, \N=8$ SCFT, which in turn is also related to
specific deformations of the $D=6$ $(0,2)$ SCFT. (The latter can
be understood because in the Bena-Warner solutions \cite{BW},
besides a stack of M2-branes we have  M5-branes with worldvolume
${\mathbb R}^{2+1}\times S^3$.) Unfortunately we do not have an
action for neither of the three and six dimensional SCFTs. Recently
there has been a proposal for obtaining an action for the $D=3$
SCFTs \cite{Schwarz}. In view of the above connections, it is an
interesting problem to check if TGMT and/or the idea of quantized
Nambu $p$-brackets can help with finding an action for either of
the two SCFTs.

\section*{Acknowledgments}
We would like to thank Seif Randjbar-Daemi for discussions. It is
a pleasure to thank the Abdus Salam ICTP, where a substantial part
of this project was done. M.M.Sh-J would like to acknowledge
partial financial support by NSF grant PHY-9870115 and funds from
the Stanford ITP.\ M.T would like to thank Center of Excellence in
Physics at the Physics department of Sharif University for its
partial financial support.
\appendix
\section{Conventions and Identities}\label{convention}
In our construction of the fuzzy four sphere, $S^4_f$, which will
be reviewed in the next Appendix, we need to develop some
identities involving $spin(5)$ gamma matrices. These are
essentially the usual four dimensional Dirac matrices which obey
the anticommutation relation, Clifford algebra \be
\l\{\gamma_i,\gamma_j\r\} = 2\delta_{ij}{\bf 1_{4\by4}} \ ,\ \ \ \
\ \ i,j=1,2,3,4.
\ee%
We follow the conventions in which the Dirac gamma matrices are
\begin{equation}\label{gamma-convention}
\gamma^{i} = \left(\begin{matrix} 0 && \sigma^i \\
\bar\sigma^i && 0 \end{matrix}\right),\ \gamma^5 =
\left(\begin{matrix} \mathbf{-1_{2\times2}} && 0 \\ 0 &&
\mathbf{1_{2\times2}} \end{matrix}\right),\ \mathbf{1_{4\times4}}
= \left(\begin{matrix} \mathbf{1_{2\times2}} && 0 \\ 0 &&
\mathbf{1_{2\times2}} \end{matrix}\right).
\end{equation}
where
\begin{equation}\label{sigma-sigma-bar}
\sigma^i = (\mathbf{1_{2\times2}}, -i\vec{\sigma}_{pauli}),\
\bar\sigma^i = (\mathbf{1_{2\times2}}, i\vec{\sigma}_{pauli})
\end{equation}
with the $\vec{\sigma}_{pauli}$ being the standard Pauli sigma
matrices:
\begin{equation}
\sigma^1 = \left(\begin{matrix}  0 && 1 \\ 1 && 0
\end{matrix}\right),\ \sigma^2 = \left(\begin{matrix}  0 && -i \\ i &&
0 \end{matrix}\right),\ \sigma^3 = \left(\begin{matrix}  1 && 0 \\
0 && -1 \end{matrix}\right),\ \mathbf{1_{2\times2}} =
\left(\begin{matrix} 1 && 0 \\ 0 && 1 \end{matrix}\right)\ .
\end{equation}
In \eqref{gamma-convention}
$\gamma^5=\gamma^1\gamma^2\gamma^3\gamma^4=\frac{1}{4!}\epsilon_{ijkl}
\gamma^i\gamma^j\gamma^k\gamma^l$.

In our conventions,
\begin{eqnarray}
 2\gamma^{ij}\equiv[\gamma^i,\gamma^j] = \left(\begin{matrix} \sigma^{ij} && 0 \nn\\
0 && \bar\sigma^{ij} \end{matrix}\right) \nn,\qquad
\gamma^i\gamma^5 = \left(\begin{matrix} 0 && -\sigma^i \nn\\
\bar\sigma^i && 0 \end{matrix}\right)
\end{eqnarray}
and also \be [\gamma^i,\gamma^j\gamma^5] = 2\delta^{ij}\gamma^5 \
, \ee where
\begin{eqnarray}
\sigma^{ij}\equiv \sigma^i\bar\sigma^j-\sigma^j\bar\sigma^i ,
&\qquad&
\bar\sigma^{ij} \equiv \bar\sigma^i\sigma^j-\bar\sigma^j\sigma^i \nn\\
\sigma^i\bar\sigma^j+\sigma^j\bar\sigma^i = 2\delta^{ij},&\qquad&
\bar\sigma^i\sigma^j+\bar\sigma^j\sigma^i = 2\delta^{ij}\ . \nn
\end{eqnarray}

$\gamma^{ij}$'s constitute a $4\by 4$ representation for the
generators of the $spin(4)$ algebra. If we add $\gamma^i\gamma^5$
to this collection, we have a representation of $spin(5)$; and the
set of $\{\gamma^{ij},\ \gamma^i\gamma^5,\ \gamma^i, \gamma^5\}$
would give the representation of the $spin(6)=su(4)$ algebra.

Next we list some useful identities involving $\sigma^i$ and
$\bar\sigma^i$:
\begin{equation}
\frac{1}{2}\epsilon^{ijkl}\sigma^{ij}
=\epsilon^{ijkl}\sigma^i\bar\sigma^j = +\sigma^{kl}\ \ ,\qquad
\frac{1}{2}\epsilon^{ijkl}\bar\sigma^{ij}=\epsilon^{ijkl}\bar\sigma^i\sigma^j
= -\bar\sigma^{kl}
\ee

\begin{eqnarray}
(\sigma^i)_{\alpha\beta}(\sigma^i)_{\lambda\rho}=
(\bar\sigma^i)_{\alpha\beta}(\bar\sigma^i)_{\lambda\rho} &=& +2
(\delta_{\alpha\beta}\delta_{\lambda\rho}-\delta_{\alpha\rho}\delta_{\beta\lambda})
\nn\\
(\sigma^i)_{\alpha\beta}(\bar\sigma^i)_{\lambda\rho} =
(\bar\sigma^i)_{\alpha\beta}(\sigma^i)_{\lambda\rho} &=&
+2\delta_{\alpha\rho}\delta_{\beta\lambda}
\end{eqnarray}
\begin{eqnarray}
(\sigma^{ij})_{\alpha\beta}(\sigma^i)_{\gamma\lambda} &=&
+2(\delta_{\gamma\lambda}\bar\sigma^j_{\alpha\beta}-
\delta_{\alpha\lambda}\bar\sigma^j_{\gamma\beta}-
\delta_{\beta\gamma}\sigma^j_{\alpha\lambda}) \nn\\
(\bar\sigma^{ij})_{\alpha\beta}(\bar\sigma^i)_{\gamma\lambda} &=&
+2(\delta_{\gamma\lambda}\sigma^j_{\alpha\beta}-
\delta_{\alpha\lambda}\sigma^j_{\gamma\beta}-
\delta_{\beta\gamma}\bar\sigma^j_{\alpha\lambda}) \\
(\sigma^{ij})_{\alpha\beta}(\bar\sigma^i)_{\gamma\lambda} &=&
-2(\delta_{\gamma\lambda}\sigma^j_{\alpha\beta}-
\delta_{\alpha\lambda}\bar\sigma^j_{\gamma\beta}-
\delta_{\beta\gamma}\sigma^j_{\alpha\lambda}) \nn\\
(\bar\sigma^{ij})_{\alpha\beta}(\sigma^i)_{\gamma\lambda} &=&
-2(\delta_{\gamma\lambda}\bar\sigma^j_{\alpha\beta}-
\delta_{\alpha\lambda}\sigma^j_{\gamma\beta}-
\delta_{\beta\gamma}\bar\sigma^j_{\alpha\lambda}) \nn
\end{eqnarray}
\begin{eqnarray}
(\sigma^{ij})_{\alpha\beta}(\sigma^{ij})_{\gamma\lambda} =
(\bar\sigma^{ij})_{\alpha\beta}(\bar\sigma^{ij})_{\gamma\lambda}&=&
16(\delta_{\alpha\beta}\delta_{\gamma\lambda}-2\delta_{\alpha\delta}
\delta_{\gamma\beta}) \\
(\sigma^{ij})_{\alpha\beta}(\bar\sigma^{ij})_{\gamma\lambda} =
(\bar\sigma^{ij})_{\alpha\beta}(\sigma^{ij})_{\gamma\lambda} &=& 0
\nn
%
\end{eqnarray}%

\section{Nambu Brackets and Fuzzy Spheres}\label{appendixB}
As fuzzy spheres are the crucial ingredients in our solutions to
the zero energy half BPS configurations of the tiny graviton
matrix theory in this appendix we present a brief review on the
fuzzy spheres. There are two equivalent approaches for the
construction of fuzzy spheres. The first, which also historically
came earlier, is based on  finding finite dimensional
representations of $SO(d+1)$ (more precisely $spin(d+1)$) for a
fuzzy $d$-sphere, $S^d_f$. This construction was first proposed
for a two sphere \cite{Madore} and then extended to four
\cite{Grosse} and to  higher dimensional spheres \cite{sunjay2}.
The second construction which were first introduced and emphasized
in \cite{tiny} is more geometrical, and is based on the
``quantization'' of the ``Nambu brackets'' \cite{Nambu}. Nambu
brackets encode the isometries of the spheres. That is this second
method which paves the way for the quantization of the giant
gravitons, which in turn leads to the matrix theory description of
the DLCQ of string/M- theory on the plane-wave backgrounds
\cite{tiny}. Hence, in \ref{Nambu-brackets} we review Nambu
brackets which are generalized form of the Poisson bracket and
prescribe a consistent way to quantize them. In
\ref{fuzzy-spheres}, using the definition of round spheres through
Nambu brackets we present the definition of fuzzy spheres and then
try to construct explicit solutions for the fuzzy two and four
sphere cases.

\subsection{Nambu brackets}\label{Nambu-brackets}
In this subsection we give the definition of Nambu brackets and
prescribe how to quantize them consistently to make multilinear
commutators.

\subsubsection*{Classical Nambu $p$-brackets}%
A Nambu $p$-bracket is defined among $p$ functions
$\F_i(\sigma^r), r,p=1,\dots,p$ as%
\be\label{Nambu-p-bracket} \l\{\F_1,\F_2,\dots,\F_p\r\} \equiv
\epsilon^{r_1r_2\cdots
r_p}\frac{\partial\F_1}{\partial\sigma^{r_1}}
\frac{\partial\F_2}{\partial\sigma^{r_p}}\dots
\frac{\partial\F_p}{\partial\sigma^{r_p}}\ ,
\ee%
where $\F_i$ are functions on a $p$-dimensional space
parameterized by $\sigma^r$. Obviously for $p=2$
\eqref{Nambu-p-bracket} reduces to the standard Poisson bracket.
These brackets, as enumerated in \cite{tiny}, have five important
properties: cyclicity, (generalized) Jacobi identity,
associativity, trace and by-part integration properties.

\subsubsection*{Quantization of Nambu $p$-brackets}

In order to pass to quantum (fuzzy, discretized or noncommutative)
physics, we generalize the standard prescription, namely start
from Hamilton-Jacobi formalism, substitute functions of phase
space coordinates with operators (matrices) acting on Hilbert
space, the Nambu brackets with multilinear
commutators (quantized Nabmu brackets) and integrals with trace, {\it i.e.}%
\begin{subequations}\label{evenpriscription}
\begin{align}
\F \longleftrightarrow \hat{\F} \qquad &;\qquad
\int d^p\sigma \longleftrightarrow Tr \\
\l\{\F_1,\cdots,\F_p\r\} \longleftrightarrow
\frac{1}{(i\hbar)^{p/2}}[\hat{\F}_1,\cdots,\hat{\F}_p]&\equiv
\frac{1}{(i\hbar)^{p/2}}\epsilon^{i_1\cdots i_{p}}
\hat{\F}_{i_1}\hat{\F}_{i_2}\cdots\hat{\F}_{i_p}
\end{align}
\end{subequations}
This prescription only works for even dimensional spaces ($p=2k$),
more precisely, for even case commutators it has   the crucial
properties of the Nambu brackets we mentioned above, that is
cyclicity, generalized Jacobi identity and by-part and trace
property. Note, however, that with (\ref{evenpriscription}b) we
have lost associativity \cite{Zachos, tiny}.

As for the odd case ($p=2k-1$), besides the associativity we also
lose the physically very important trace and by-part integration
properties. To overcome that problem, in \cite{tiny} it was
proposed to replace an odd $p$-bracket with an even
$(p+1)$-bracket and then again use (\ref{evenpriscription}b). In
order that, we should introduce a fixed matrix, a non-dynamical
operator, $\Lodd$ (at least in string theory and in the case of
fuzzy odd spheres we have strong evidence where it comes from,
e.g. see discussions of \cite{tiny} and also section
\ref{mass-deformed}).
The prescription for quantization of odd brackets is then%
\be\label{oddpriscription}%
 \F \longleftrightarrow \hat{\F}
\ ;\ \l\{\F_1,\cdots,\F_{2k-1}\r\} \longleftrightarrow
\frac{1}{(i\hbar)^{k}}\l[\hat{\F}_1,\cdots,\hat{\F}_{2k-1},\Lodd\r]
\ ;\  \int d^{2k-1}\sigma \longleftrightarrow Tr\ .
\ee%

\subsection{Fuzzy spheres}\label{fuzzy-spheres}
Any round (classical, continuous, commutative) $d$-sphere with
radius $R$
should satisfy these constraints as defining properties%
\be\label{classical-sphere-def}
 \sum _{\mu=1}^{d+1}(X^\mu)^2 = R^2 \qquad;\qquad
\{X^{\mu_1},X^{\mu_2},\cdots,X^{\mu_d}\} =
R^{d-1}\epsilon^{\mu_1\cdots
\mu_{d+1}}X^{\mu_{d+1}}. \ee%
The $d+1$ embedding coordinates $X^{\mu}$ rotate as a vector under
the isometry symmetry of the $d$-sphere, $SO(d+1)$. In fact the
bracket equation in \eqref{classical-sphere-def} is a
manifestation of the $SO(d+1)$ isometry. (Note that $SO(d+1)$ has
only two invariant tensors, the metric $\delta_{\mu\nu}$ and
$\epsilon^{\mu_1\cdots \mu_{d+1}}$.)

In order to pass to the fuzzy (discretized or noncommutative)
$d$-sphere, we follow the prescription for quantizing Nambu
brackets presented above. Let us first start with a fuzzy even
sphere. A fuzzy $2k$ sphere, $S^{2k}_f$, is defined through
\be\label{fuzzyevensphere} \sum_{\mu=1}^{2k+1}(X^\mu)^2 = R^2
\qquad;\qquad \l[X^{\mu_1},X^{\mu_2},\cdots,X^{\mu_{2k}}\r] =
i^k\lambda^{2k-1}\epsilon^{\mu_1\mu_2\cdots\mu_{2k+1}}X^{\mu_{2k+1}}, %
\ee%
which leads to
\[
\epsilon^{i_1i_2\cdots i_{2k+1}}X^{i_1}X^{i_2}\cdots X^{i_{2k+1}}
= i^k\lambda^{2k-1}R^2.
\]
In the above, we have defined the ``${\hbar}$'' ({\it cf.}
(\ref{evenpriscription}b)) as $(\lambda/R)^{(2k-1)/k}$ and
$\lambda$, which in a sense quantifies how much our sphere is far
from the classical sphere, will be called the ``fuzziness''.

Similarly for the fuzzy odd-spheres%
\be\label{fuzzyoddsphere} \sum_{i=1}^{2k}(X^i)^2 = R^2
\qquad;\qquad \l[X^{i_1},X^{i_2},\cdots,X^{i_{2k-1}},\Lodd\r] =
i^kl^{2(k-1)}\epsilon^{i_1i_2\cdots i_{2k}}X^{i_{2k}} %
\ee%
leading to
\[
 \epsilon^{i_1i_2\cdots i_{2k+1}}X^{i_1}X^{i_2}\cdots\Lodd
X^{i_{2k}} = i^kl^{2(k-1)}R^2.
\]

With the above definition, the problem of construction of fuzzy
even spheres is now reduced to finding explicit solutions to
\eqref{fuzzyevensphere} in terms of some $N\by N$ matrices and
giving the relation between the size of the sphere (its radius)
and the size of the matrices. For the fuzzy odd sphere case, in
addition we should also give an appropriate $\Lodd$. Note that by
construction, solutions to \eqref{fuzzyevensphere} or
\eqref{fuzzyoddsphere} are only specified up to an $SO(d+1)$
rotation:
\[
{\cal R}_{ij}X^j=U({\cal R}) X_i U({\cal R})^{-1},
\]
where ${\cal R}_{ij}$ are the generators of the $spin(d+1)$
isometry group of the sphere and $U({\cal R})$ are an $N\by N$
representation of ${\cal R}$.

In this appendix, as examples, we will construct explicit
solutions to \eqref{fuzzyevensphere} for $k=1,2$ cases, {\it i.e.}
fuzzy two and four spheres. The construction of $S^3_f$ is
presented in section \ref{Zero-Energy-section} of the main text.
There are two ways to solve these equations and realize the fuzzy
spheres, group theoretical and harmonic oscillator approaches. The
first approach is widely studied and used since the introduction
of fuzzy spheres was based on the group and the representation
theory of the isometry of the sphere under study. The harmonic
oscillator approach is a less-known one and is well-suited for
specific dimension, and is inspired by principal Hopf fibrations.
This method for the fuzzy two sphere were studied in some detail
in \cite{Hammou}.

\subsubsection{ Group theoretical construction
of fuzzy spheres}\label{group-theory-construct}%
The fuzzy $2k$-sphere is defined through the algebra $\M_N$ of
$N\by N$ hermitian matrices. When $N\rightarrow\infty$ the
corresponding matrix geometry tends to the geometry of the round
$2k$-sphere. To realize this idea, we think of the ordinary
geometry of the round sphere as an infinite dimensional
representation of $spin(2k+1)$ algebra. To pass to the fuzzy case,
however, we need finite dimensional representations. Due to the
basic theorem of the algebraic geometry, one can replace a given
manifold with the algebra of functions defined on that. For the
case of the $S^{2k}$ that is the algebra $\F$ of functions
$f(X^\mu)$ on $S^{2k}$ which admits a polynomial expansion in the $X^\mu$'s:%
\be%
f(X_\mu) = f_0 + f_\mu X^\mu + \frac{1}{2}f_{\mu\nu}X^\mu X^\nu +
\frac{1}{6}f_{\mu\nu\rho}X^\mu X^\nu X^\rho + \cdots%
\ee%
Now we would like to {\it truncate} this expansion in such a way
that it is convertible to an algebra. To turn this approximation
to algebra we replace $X^\mu$ with generalized gamma matrices
$\G^\mu_{2k}$ of the algebra $\M_{N_k(n)}$ of $N_k(n)\by N_k(n)$
hermitian matrices. These matrices are defined as follows%
\bea\label{G-general} %
\G^{\mu} &=&
(\Gamma^\mu\otimes\mathbf1\otimes\cdots\otimes\mathbf1 +
\mathbf1\otimes\Gamma^\mu\cdots\otimes\mathbf1 +
\cdots+\mathbf1\otimes\cdots\mathbf1\otimes\Gamma^\mu)_{Sym} \cr
\G^{\mu\nu} &=&
(\Gamma^{\mu\nu}\otimes\mathbf1\otimes\cdots\otimes\mathbf1 +
\mathbf1\otimes\Gamma^{\mu\nu}\cdots\otimes\mathbf1 +
\cdots+\mathbf1\otimes\cdots\mathbf1\otimes\Gamma_{\mu\nu})_{Sym}%
\eea%
where $\G^{\mu\nu} = \frac{1}{2}\l[\G^\mu,\G^\nu\r]$ and
$\Gamma^{\mu\nu} =\frac{1}{2}\l[\Gamma^\mu,\Gamma^\nu\r] =
\Gamma^\mu\Gamma^\nu$ are the generators of $SO(2k+1)$. These
generalized gamma matrices are $n$-fold direct tensor-product of
gamma matrices and identity matrix and the symmetrization means
that they are defined on the vector space $Sym\l(V^{\otimes
n}\r)$, the symmetrized $n$-fold tensor product space of the
smallest irreducible representation of $spin(2k+1)$, $V$; and
$dim(Sym\l(V^{\otimes n}\r))=N(n)$. Hence, $\G^{\mu}$ are $N\by N$
matrices.
These matrices have following properties%
\begin{subequations}\label{radius-n}
\begin{align}
 \sum_{\mu=1}^{2k+1} \G_\mu\G_\mu &= n(n+2k)\ ,\\
\l[\G^{\mu_1},\G^{\mu_2},\dots,\G^{\mu_{2k}}\r] &= -(-i)^k
(2k)!!(2k-2+n)!!\epsilon^{\mu_1\mu_2\cdots
\mu_{2k+1}}\G^{\mu_{2k+1}}.
\end{align}
\end{subequations}%
If we now define  $X_\mu = \tilde{\lambda}\G^{(2k)}_\mu$ we can
see that this matrix construction truly gives $S_f^{2k}$. The only
remaining step to complete the construction is then to specify $N$
as a function of $n$, which in turn, noting \eqref{radius-n},
gives the relation between the size of matrices $N$ and the radius
of the fuzzy sphere $R$. We will work this out for two specific
cases of $S^2_f$ and $S^4_f$.
\newline%
{\bf Construction of $S_f^2$}

Here we deal with $spin(3)=SU(2)$, and hence the analog of
\eqref{G-general} for this case is the
generalized Pauli sigma matrices,%
\be %
\Sigma_i = (\sigma_i \otimes\mathbf1\otimes\cdots\otimes\mathbf1 +
\mathbf1\otimes\sigma_i\otimes\cdots\otimes\mathbf1 +
\cdots+\mathbf1\otimes\cdots\otimes\mathbf1\otimes\sigma_i)_{Sym}\
,
\ee%
where $\sigma^i$ are $2\by2$ Pauli matrices. $\Sigma_i$ form an
$(n+1)\by(n+1)$ representation of the $SU(2)$ algebra,
corresponding to spin $n/2$ states.
Note that for $SU(2)$%
\be\label{Nvs.n-SU(2)} N\equiv dim(Sym(V^{\otimes}))=n+1\ . \ee
The embedding coordinates can be identified as $X^i =
\frac{\lambda}{2}\Sigma^i$. The size of the matrices and the
radius of the sphere are then related as \be
R^2_{S^2_f}=\frac{\lambda^2}{4}(N^2-1). \ee
\newline%
{\bf Construction of $S_f^4$}

Here the basic $\gamma$-matrices we start with are the standard
$4\by 4$ Dirac $\gamma$-matrices, together with $\gamma^5$, for
our conventions see Appendix \ref{convention}. These $\gamma$
matrices satisfy
\[
\sum_{\mu=1}^5 \gamma^\mu\gamma^\mu = 5 \qquad;\qquad
[\gamma^\mu,\gamma^\nu,\gamma^\rho,\gamma^\alpha] = 4!
\epsilon^{\mu\nu\rho\alpha\beta} \gamma^\beta
\]%
Again one can use \eqref{G-general} to construct an $N\by N$
representation out of these $\gamma$'s. The only remaining part is
how $N$ and the radius of the fuzzy four sphere are related. To
work this out, one should use representation theory of $spin(5)$.
This has been carried out in \cite{sunjay2} and the result is
$N=\frac{1}{6}(n+1)(n+2)(n+3)$, where $R^2_{S^4_f}\propto n(n+4)$.

\subsubsection{Harmonic oscillator construction
of fuzzy spheres}\label{harmonic-oscil-construct}%

Commutative $d$-spheres are usually defined through embedding the
$S^d$ in ${\mathbb R}^{d+1}$. The natural question to ask is
whether a similar thing is also possible for a noncommutative
fuzzy sphere. Precisely, the question is whether it is possible to
embed an $S^d_f$ into a noncommutative plane, preferably a
noncommutative Moyal plane. The answer, if positive, cannot be a
$d+1$ dimensional plane. This can be seen from the fact that
imposing the noncommutativity on the coordinates of the plane
reduces the $SO(d+1)$ rotational invariance. Hence, we should look
for embedding the $S^d_f$ into a higher dimensional Moyal plane,
which includes $SO(d+1)$ among its isometries. In working with
Moyal plane, it is more convenient to adopt  complex coordinates.
Let us consider a $\C^p$ space parameterized with the coordinates
$z_\alpha,\ \alpha=1,2,\cdots, p$ which satisfy the following
commutation relation \be\label{Moyal-def}
 [z_\alpha,\bar z_\beta] = \theta\delta_{\alpha\beta}\ .
\ee%
We  denote this space by $\C^p_\theta$. (In terms of
$a_\alpha=z_\alpha/\sqrt{\theta}$, the above is nothing but the
algebra of a $p$ dimensional harmonic oscillator.) The idea is to
solve \eqref{fuzzyevensphere} for an $S^{2k}_f$ using
\eqref{Moyal-def}. The first thing to specify is what is the
appropriate $p$ for a given $k$. For that we note that
$\C^p_\theta$, and specifically \eqref{Moyal-def}, is invariant
under $U(p)$ transformations which rotate $z_\alpha$ to $(U\
z)_{\alpha}$, $U\in U(p)$. Therefore, as the first criterion,
$SO(2k+1)$ should be a subgroup of $U(p)$, e.g. for $k=1$, the
case of a two sphere, $p$ can be two and for $k=2$, the four
sphere case, $p$ should at least be four. For a generic case, one
may start with
\be\label{X-z}%
 X^\mu =\kappa\bar z_\alpha\l(\Gamma^\mu\r)_{\alpha\beta}z_\beta\ ,
\ee%
where $\Gamma^\mu$ are $2k+1$ dimensional Dirac matrices, and
hence are $D\by D$ matrices, where $D$ is the dimension of the
smallest fermionic representation of $SO(2k+1)$. As is evident
from \eqref{X-z} the proper choice is $p=D$. For $k=1$ and $2$,
$D=2^{2/2}=2$ and $2^{4/2}=4$. In \eqref{X-z} $\kappa$ is a
parameter of dimension one over length.

We would like to use \eqref{X-z} as the relations embedding a
sphere in Moyal plane. In \eqref{X-z} the $\Gamma^\mu$ matrices
are essentially Clebsch-Gordon coefficients relating $U(p)$
vectors to $SO(2k+1)$ vectors. It is worth-noting that $X^\mu$'s
defined in \eqref{X-z}, similarly to the $\Gamma^\mu$'s, are
hermitian and well-suited for our purpose.

As a warm-up let us first focus on the case of the fuzzy two
sphere and then discuss the case of four sphere and possible
generalizations to higher dimensional spheres.
\newline%
{\bf Construction of $S^2_f$}

We are looking for an embedding of an $S^2_f$ in a $C^2_\theta$.
This is done in \cite{Hammou} and here we briefly review that. To
start with, consider \be\label{S2-C2} X^i=\frac{\kappa}{2} \bar
z_\alpha (\sigma^i)_{\alpha\beta} z_\beta\ , \ee where $i=1,2,3$
and $\alpha,\beta=1,2$. If $[z_\alpha,\bar
z_\beta]=\theta\delta_{\alpha\beta}$, it is straightforward to
show that \be\label{SU(2)-z} [X^i, X^j]=i\kappa\theta\
\epsilon_{ijk}X_k\ . \ee Furthermore, one can easily show that
\be\label{two-sphere-radius} \sum_{i=1}^3 (X^i)^2=
\kappa^2\theta^2\ X^0 (X^0+\bf1) , \ee where we have defined
\[
X^0=\frac{1}{2\theta} \bar z_\alpha z_\alpha\ .
\]
{}From \eqref{SU(2)-z} the fuzziness, $\lambda$, is identified as
\be\label{lambda-S2} \lambda=\kappa\theta\ , \ee and hence \be
R^2_{S^2_f}= \lambda^2 X^0 (X^0+1) . \ee It is noteworthy that
$[X^0, f(X^i)]=0$, where $f$ is a generic function of $X^i$'s. If
we choose $\kappa$, which is so far a free parameter, to be
$1/\sqrt{\theta}$ then $\lambda=\sqrt{\theta}$. That is, the fuzziness
is equal to the noncommutativity parameter of the embedding Moyal
plane. (Note that the minimal area which one can measure in the
Moyal plane is $\theta$.) Another interesting choice for $\kappa$
can be $1/R$. We will return to this choice and its physical
significance later.

$X^i$ and $X^0$, as matrices are infinite dimensional, however, we
are looking for finite dimensional matrices. In addition, in order
to have a fuzzy two sphere (of a given radius) we need $\sum
(X^i)^2$ to be proportional to the identity matrix. Noting
\eqref{two-sphere-radius}, this means that we should restrict
$X^0$ to a block in which it is proportional to identity. This can
be easily done, recalling that $X^0$ is the number operator for a
two dimensional harmonic oscillator and hence in the number
operator basis it takes a diagonal form, consisting of blocks
$(n+1)\by (n+1)$ blocks with the eigenvalue $n$. (For a two
dimensional harmonic oscillator the multiplicity of a state of
energy $n$ is $n+1$.) Therefore, if we focus only on
$X^0=\frac{1}{2}n$ sector, we have a description of a fuzzy two
sphere with radius $R^2_{S^2_f}=\lambda^2
\frac{n}{2}(\frac{n}{2}+1)$, in terms of $(n+1)\by (n+1)$
matrices.

It is instructive to study some interesting limits of the above
fuzzy two sphere. In our problem we have two parameters, the
fuzziness $\lambda$, and the radius of the sphere $R$. One may
study various limits keeping some combinations of the two fixed.
For example, the commutative $S^2$ limit is when $\lambda\to 0$,
keeping $R$ fixed. (In this limit it is more convenient to choose
the normalization $\kappa\sim 1/R$.) Another interesting limit is
the Moyal plane limit \cite{Chu-Madore}: \be\label{Moyal-plane}
\lambda\to 0\ ,\quad R\to\infty\ ;\qquad \lambda R=fixed\ . \ee
Intuitively one can think of this limit by choosing one of the
$X$'s, say $X^3$ to be very close to $R$ and $X^1,X^2\ll R$. In
this limit the commutation relation \eqref{SU(2)-z} would then
reduce to
\[
[X^1, X^2]=i \lambda R\ ,
\]
which defines a two dimensional Moyal plane. The minimal area that
one can measure on this sphere is $\lambda R$. This result is in
fact more general than this limit and is true for a generic
$S^2_f$ of finite radius. That is, $\lambda R$, and not
$\lambda^2$,  is the minimal area which can be measured on a fuzzy
two sphere.

Before moving to the $S^4_f$ example, we would like to comment
more on the relation between the $U(2)$ symmetry of the embedding
$\C^2_\theta$ space and $spin(3)=SU(2)$ of the $S^2_f$. The extra
$U(1)$ is basically the transformations which rotate $z_1,\ z_2$
in the same way. The $X^i$'s, by construction, are explicitly
invariant under this $U(1)$. The generator of this $U(1)$ is
$X^0$. Geometrically the embedding \eqref{S2-C2} is a realization
of the Hopf fibration. For the moment let us consider the
commutative $\theta=0$ case. The $X^0=const.$ surface then defines
an $S^3$ in ${\mathbb R}^4$. The $S^3$, however, can be thought as
an $S^1$ fiber over an $S^2$ base (e.g. see \cite{Hammou} and
references therein). To reduce $S^3$ to $S^2$ we then need to
reduce over the $S^1$ fiber. This is done in \eqref{S2-C2} by
taking the $X^i$ which are invariant under the global $U(1)$
rotating $z$'s with the same phase.
\newline%
{\bf Construction of $S^4_f$}

As we argued for the $S^4_f$ case the embedding space should be
$\C^4_\theta$, as four dimensional Dirac $\gamma$-matrices are
$4\by 4$. For performing computations it turns out to be more
convenient if parameterize $\C^4_\theta$ as two $\C^2_\theta$'s,
with $u_\alpha$ and $v_\alpha$, $\alpha=1,2$ coordinates, where
\be\label{u-v-NC}%
 [u_\alpha,\bar u_\beta] =
\theta\delta_{\alpha\beta} \quad;\quad [u_\alpha,\bar v_\beta] = 0
\quad ;\quad [v_\alpha,\bar v_\beta] = \theta\delta_{\alpha\beta}\ .%
\ee%
We can again use \eqref{X-z}, to obtain the embedding coordinates.
Employing the conventions of Appendix \ref{convention} for the
Dirac matrices, we
have%
\begin{subequations}\label{S4-embedding}%
\begin{align}
X^i&=\kappa (\bar u \sigma^i v+ \bar v \bar\sigma^i u)\ , \qquad i=1,2,3,4\\
X^5&=\kappa (\bar v {\bf 1} v- \bar u {\bf 1} u)\ , \\
X^0&= \frac{1}{\theta}(\bar u {\bf1}u + \bar v{\bf1}v)\ ,
\end{align}
\end{subequations}%
where $\sigma^i$'s are defined in \eqref{sigma-sigma-bar}. In the
``classical'' $\theta=0$ case \eqref{S4-embedding} is a
realization of the Hopf map for S$^7$ with an $S^4$ base.

It is straightforward, but perhaps tedious, to show that the
embedding coordinates \eqref{S4-embedding} satisfy%
\bea%
\sum _{\mu=1}^5(X^\mu)^2 &=& \kappa^2 \theta^2 X^0 (X^0+ 4\cdot\bf 1)\\
\l[X^\mu,X^\nu,X^\rho,X^\alpha\r] &=& \frac{1}{3}\kappa^3\theta^3
(X^0+2\cdot{\bf 1})\ \epsilon^{\mu\nu\rho\alpha\beta}X^\beta.
\label{four-bracket-X}\eea%
Therefore, if we restrict $X^0$ to a block in which it is
proportional to the identity matrix, \eqref{S4-embedding} is
defining a fuzzy four sphere with the radius $R_{S^4_f}$ and the
fuzziness
\be\label{S4-fuzziness}%
\lambda^3=\frac{1}{3}(\kappa\theta)^3\left(\frac{R^2_{S^4_f}}
{(\kappa\theta)^2}-4\right)
\ee%
For more details see the main text, section
\ref{single-giant-section}.

As in the case of the $S^2_f$ one can study some interesting
limits, e.g. the commutative round $S^4$ limit is obtained when
$\lambda\to 0$, keeping the radius fixed. The other interesting
limit is the four dimensional noncommutative plane limit, {\it
i.e.} $\lambda\to 0,\ R\to\infty$ keeping $\lambda^3 R$ fixed. One
can think of this limit as expansion of the four sphere about its
north pole, {\it i.e.} take one of the $X^\mu$'s, say $X^5$, to be
very close to $R$ while $X^i\ll R$. In this limit
\eqref{four-bracket-X} reduces to %
\be\label{L4}%
 [X^i,X^j, X^k, X^l]=\lambda^3 R\ \epsilon^{ijkl}\ \equiv L^4 \
 \epsilon^{ijkl}.%
\ee%
Note that this noncommutative plane is {\it not} a Moyal plane. As
it can be seen from \eqref{L4}
\be\label{L4=l2R2}%
L^4=\lambda^3 R=l^2 R^2%
\ee%
(and not $\lambda^4$) is the smallest volume on this plane that
one can measure. To obtain the second equality in \eqref{L4=l2R2}
we have used \eqref{normalizations} and \eqref{S3f-radius}. This
result can be shown to be true beyond this limit, for an $S^4_f$
of generic radius.

One appropriate and natural choice for the normalization
coefficient $\kappa$ is then obtained when we identify $L$ with
the noncommutativity scale of the embedding eight dimensional
Moyal plane, {\it i.e.}%
\be\label{theta-L2}%
L^2=\theta\ . %
\ee%
 This leads to the normalization \eqref{normalization-choices}.
\section{Superalgebra of the Tiny Graviton Matrix Theory}\label{superalgebra}
The plane-wave \eqref{background} is a maximally supersymmetric
one, {\it i.e.} it has 32 fermionic isometries which can be
arranged into two sets of 16, the kinematical supercharges, $q$'s,
and the dynamical supercharges, $Q$'s. The former are those which
anticommute to light-cone momentum $P^+$ and the latter
anticommute to the light-cone Hamiltonian ${\bf H}$. Here we show
the dynamical part of superalgebra, which can be identified with
\super \ and adopt the conventions of \cite{review}. For the
complete superalgebra see \cite{review}. \bea [P^{+},
q_{\alpha\beta}]=0 \quad&,&\quad [P^{+},
q_{\dot\alpha\dot\beta}]=0\ ,\cr [{\bf H}, q_{\alpha\beta}]=-i\mu
q_{\alpha\beta} \quad&,&\quad [{\bf H},
q_{\dot\alpha\dot\beta}]=i\mu q_{\dot\alpha\dot\beta}\ . \eea \bea
[P^{+}, Q_{\alpha\dot\beta}]=0 \quad&,&\quad [P^{+},
Q_{\dot\alpha\beta}]=0 \cr [{\bf H}, Q_{\alpha\dot\beta}]=0
\quad&,&\quad [{\bf H},Q_{\dot\alpha\beta}]=0
\eea%
\be\label{qq} \{q_{\alpha
\beta},q^{\dagger\rho\lambda}\}=2P^+\delta_{\alpha}^{\ \rho}
\delta_{\beta}^{\ \lambda}\  , \ \ \ \{q_{\alpha
\beta},q^{\dagger\dot\alpha \dot\beta}\}=0\ ,\ \ \{q_{\dot\alpha
\dot\beta},q^{\dagger\dot\rho
\dot\lambda}\}=2P^+\delta_{\dot\alpha}^{\
\dot\rho}\delta_{\dot\beta}^{\ \dot\lambda}\  ,%
\ee%
\bea\label{QQ}%
 \{Q_{\alpha\dot\beta},Q^{\dagger\rho
\dot\lambda}\}&=&2\ \delta_{\alpha}^{\ \rho} \delta_{\dot\beta}^{\
\dot\lambda}\ {\bf H} + \mu (i\sigma^{ij})_{\alpha}^{\ \rho}
\delta_{\dot\beta}^{\ \dot\lambda}\ {\bf J}^{ij} + \mu
(i\sigma^{ab})_{\dot\beta}^{\ \dot\lambda}\delta_{\alpha}^{\ \rho}
{\bf J}^{ab} \ , \cr
\{Q_{\alpha \dot\beta},Q^{\dagger\dot\rho \lambda}\}&=& 0 \ , \\
\{Q_{\dot\alpha \beta},Q^{\dagger\dot\rho \lambda}\}&=&2\
\delta_{\dot\alpha}^{\ \dot\rho} \delta_{\beta}^{\ \lambda}\ {\bf
H} +\mu (i\sigma^{ij})_{\dot\alpha}^{\ \dot\rho} \delta_{\beta}^{\
\lambda}\ {\bf J}^{ij} + \mu (i\sigma^{ab})_{\beta}^{\
\lambda}\delta_{\dot\alpha}^{\ \dot\rho} {\bf J}^{ab} \ .
\nonumber%
\eea%

The generators of the above supersymmetry algebra can be realized
in terms of $J\by J$ matrices as
\begin{eqnarray}
P^+=-P_- = \frac{1}{R_-} \Tr {\bf 1} \qquad&,&\qquad
P^-=-P_+=-{\bf H} \cr q_{\alpha\beta}=\frac{1}{\sqrt{R_-}}\ \Tr
\psi_{\alpha\beta}\qquad&,&\qquad
q_{\dot\alpha\dot\beta}=\frac{1}{\sqrt{R_-}}\ \Tr
\psi_{\dot\alpha\dot\beta}
\end{eqnarray}
\begin{eqnarray}
{\bf J}_{ij} &=& \Tr \left(X^i\Pi^j-X^j\Pi^i +
\psi^{\dagger\alpha\beta} (i\sigma^{ij})_{\alpha}^{\ \rho}
\psi_{\rho\beta} -\psi^{\dagger \dot\alpha\dot\beta}
(i\sigma^{ij})_{\dot\alpha}^{\ \dot\rho}
\psi_{\dot\rho\dot\beta}\right) \cr {\bf J}_{ab} &=& \Tr
\left(X^a\Pi^b-X^b\Pi^a + \psi^{\dagger \alpha\beta}
(i\sigma^{ab})_{\beta}^{\ \rho} \psi_{\alpha\rho} -\psi^{\dagger
\dot\alpha\dot\beta} (i\sigma^{ab})_{\dot\beta}^{\ \dot\rho}
\psi_{\dot\alpha\dot\rho}\right)
\end{eqnarray}
\be\begin{split} Q_{\dot\alpha\beta}=\sqrt{\frac{R_-}{2}}\ \Tr
&\Bigl[ (\Pi^i-i\frac{\mu}{R_-} X^i) (\sigma^{i})_{\dot\alpha}^{\
\rho} \psi_{\rho\beta}+ (\Pi^a-i\frac{\mu}{R_-} X^a)
(\sigma^{a})_{\beta}^{\ \dot\rho} \psi_{\dot\alpha\dot\rho}\cr +&
\frac{1}{3! g_s} \left(\epsilon^{ijkl}[ X^i, X^j, X^k, {\cal L}_5]
(\sigma^{l})_{\dot\alpha}^{\ \rho}\psi_{\rho\beta}+
\epsilon^{abcd}[ X^a, X^b, X^c, {\cal L}_5]
(\sigma^{d})_{\beta}^{\
\dot\rho}\psi_{\dot\alpha\dot\rho}\right)\cr +& \frac{1}{2
g_s}\left( [ X^i, X^a, X^b, {\cal L}_5]
(\sigma^{i})_{\dot\alpha}^{\ \rho} (i\sigma^{ab})_{\beta}^{\
\gamma}\psi_{\rho\gamma}+ [ X^i, X^j, X^a, {\cal L}_5]
(i\sigma^{ij})_{\dot\alpha}^{\ \dot\rho} (\sigma^{a})_{\beta}^{\
\dot\gamma}\psi_{\dot\rho\dot\gamma}\right)\Bigr]
\end{split}
\ee%
\be\begin{split} Q_{\alpha\dot\beta}=\sqrt{\frac{R_-}{2}}\ \Tr
&\Bigl[ (\Pi^i-i\frac{\mu}{R_-} X^i) (\sigma^{i})_{\dot\alpha}^{\
\rho} \psi_{\rho\beta}+ (\Pi^a-i\frac{\mu}{R_-} X^a)
(\sigma^{a})_{\beta}^{\ \dot\rho} \psi_{\dot\alpha\dot\rho}\cr +&
\frac{1}{3! g_s} \left(\epsilon^{ijkl}[ X^i, X^j, X^k, {\cal L}_5]
(\sigma^{l})_{\dot\alpha}^{\ \rho}\psi_{\rho\beta}+
\epsilon^{abcd}[ X^a, X^b, X^c, {\cal L}_5]
(\sigma^{d})_{\beta}^{\
\dot\rho}\psi_{\dot\alpha\dot\rho}\right)\cr +& \frac{1}{2
g_s}\left( [ X^i, X^a, X^b, {\cal L}_5]
(\sigma^{i})_{\dot\alpha}^{\ \rho} (i\sigma^{ab})_{\beta}^{\
\gamma}\psi_{\rho\gamma}+ [ X^i, X^j, X^a, {\cal L}_5]
(i\sigma^{ij})_{\dot\alpha}^{\ \dot\rho} (\sigma^{a})_{\beta}^{\
\dot\gamma}\psi_{\dot\rho\dot\gamma}\right)\Bigr]
\end{split}\ee%

The (anti)commutation relations may be verified using the quantum
(as opposed to matrix) commutation relations:%
\bea [X^I_{pq}, \Pi^J_{rs}] &=& i\delta^{IJ}\
\delta_{ps}\delta_{qr} \cr \{(\psi^{\dagger \alpha\beta})_{pq},
(\psi_{\rho\gamma})_{rs}\} &=& \delta^{\alpha}_{\rho}
\delta^{\beta}_{\gamma}\ \delta_{ps}\delta_{qr} \cr
\{(\psi^{\dagger \dot\alpha\dot\beta})_{pq},
(\psi_{\rho\gamma})_{rs}\} &=& \delta^{\dot\alpha}_{\dot\rho}
\delta^{\dot\beta}_{\dot\gamma}\ \delta_{ps}\delta_{qr}
\eea%
where $p,q, r, s=1,2,\cdots , J$ are matrix indices.%
%

\end{document}